\RequirePackage{fix-cm}
\documentclass{svjour3}                     % onecolumn (standard format)
\smartqed  % flush right qed marks, e.g. at end of proof
\usepackage{url}
\usepackage{graphicx}
\usepackage{subfig}
\journalname{SNAM}
\begin{document}

\title{Development of Computer Science Disciplines - A Social Network Analysis Approach}

\author{Manh Cuong Pham \and
 Ralf Klamma \and
 Matthias Jarke}
 \authorrunning{M.C. Pham, R.Klamma, M.Jarke}
 \institute{Manh Cuong Pham \at
              Information Systems and Database Technology\\
              RWTH Aachen University, Aachen\\
              Ahornstr. 55, D-52056, Aachen, Germany\\
              \email{pham@dbis.rwth-aachen.de}
           \and
           Ralf Klamma \at
               Information Systems \& Database Technology\\
               RWTH Aachen University, Aachen\\
               Ahornstr. 55, D-52056, Aachen, Germany\\
               \email{klamma@dbis.rwth-aachen.de}
           \and
           Matthias Jarke \at
              Information Systems \& Database Technology\\
              RWTH Aachen University, Aachen\\
              Ahornstr. 55, D-52056, Aachen, Germany\\
              \email{jarke@dbis.rwth-aachen.de}
}

\maketitle
\thispagestyle{empty}

\begin{abstract}
% How do the disciplines in computer science develop over time? Why some journals or conferences are so successful, while some others are not?
   In contrast to many other scientific disciplines, computer science considers conference publications. Conferences have the advantage of providing fast publication of papers and of bringing researchers together to present and discuss the paper with peers. Previous work on knowledge mapping focused on the map of all sciences or a particular domain based on ISI published JCR (Journal Citation Report). Although this data covers most of important journals, it lacks computer science conference and workshop proceedings. That results in an imprecise and incomplete analysis of the computer science knowledge. This paper presents an analysis on the computer science knowledge network constructed from all types of publications, aiming at providing a complete view of computer science research. Based on the combination of two important digital libraries (DBLP and CiteSeerX), we study the knowledge network created at journal/conference level using citation linkage, to identify the development of sub-disciplines.  We investigate the collaborative and citation behavior of journals/conferences by analyzing the properties of their co-authorship and citation subgraphs. The paper draws several important conclusions. First, conferences constitute social structures that shape the computer science knowledge. Second, computer science is becoming more interdisciplinary. Third, experts are the key success factor for sustainability of journals/conferences.
\end{abstract}

%\keywords{citation analysis, knowledge network, social network, digital library}

%-------------------------------------------------------------------------
\section{Introduction}

Recent studies on knowledge mapping in scientometrics are concerned with building, visualizing and qualitatively analyzing the knowledge networks of sciences \cite{Boyack05,Boyack07,leydesdorff04,Moya04}. Similar to the geographical map, the knowledge network of sciences, or the map of sciences is used to provide us an insight into the structure of science. It can be used to visually identify major areas of science, their similarity and interconnectedness. Methods developed in bibliometrics and scientometrics such as citation analysis, content analysis and recently proposed method based on clickstream data \cite{Bollen09} are commonly used in this domain.

Computer science is a fast-changing research field. Unlike other disciplines where academic standard of publishing is to publish in journals, in computer science we consider conference publication. Previous work on knowledge mapping typically focused on single disciplines \cite{Tsay03,Ding00,McCain98,Boyack07} or on the whole science \cite{Boyack05,Klavans06,Bollen09} based on the analysis of massive citation data such as Journal Citation Report (JCR), Science Citation Index (SCI), Science Citation Index Expanded (SCIE) and Social Science Citation Index (SSCI), published by Thompson Scientific (TS, formally ISI). Those datasets cover most of important journals of science, but they do not contain computer science conference and workshop proceedings. That makes any attempt to map computer science knowledge either imprecise or limited to small fields.

With the recent availability of large-scale citation index from digital libraries in computer science such as ACM Portal\footnote{\url{http://portal.acm.org/portal.cfm}}, IEEE Xplore\footnote{\url{http://ieeexplore.ieee.org/Xplore/dynhome.jsp}}, DBLP\footnote{\url{http://www.informatik.uni-trier.de/~ley/db/}} and CiteSeerX\footnote{\url{http://citeseerx.ist.psu.edu/}}, it is possible to study the relationship between publication venues and provide a more precise and complete view of today's computer science research landscape at both local and global scale. In this paper (some of results are published in an earlier conference paper \cite{Pham10}), we are concerned with studying the structure of knowledge network and the publication culture in computer science. Using the combination of two large important digital libraries in computer science, DBLP and CiteSeerX, we build a so-called knowledge map of the computer science and provide a comprehensive visualization which allows us to explore its macro structure and its development over time. To get an insight into the collaborative and citation behavior in computer science, we investigate the graphical features of the citation and collaboration subgraphs of journals/conferences. One of our main findings is that conferences constitute social structures that shape the computer science knowledge. By analyzing the combined knowledge network of journal and conference publications, we are able to identify clusters (or sub-disciplines) and trace their development, which is not possible by the analysis of journals only. We also find that computer science publications are very heterogeneous and the field is becoming more interdisciplinary as each sub-disciplines tends to connect to many other sub-disciplines. Finally, there is a connection between the local structure of the citation and collaboration subgraphs of journals/conferences and their impact. On the one hand, high impact journals/conferences successfully build the core topic and attract the contributions from research community. On the other hand, experts are the key success factor for maintaining and cultivating the community of journals/conferences (hereafter called venues).

The paper is organized as follows. In Section 2, we briefly survey the related work. In Section 3, we discuss about the role of conferences in computer science. In Section 4, we describe the data set used in our study. In Section 5, the creation of networks used in our study is presented. In Section 6, we discus about the network visualization. In Section 7 we discuss about the development of sub-disciplines in computer science. In Section 8, we present the venues ranking using SNA metrics. In section 9, we present our analysis on the properties of venue's subgraph and their relation to the impact of venues. The paper finishes with some conclusions and our directions for future research.

%-------------------------------------------------------------------------
\section{Related Work}
Social network analysis and visual analytics have been applied to represent the knowledge \cite{Leyla10}, to detect the communities and hierarchical structures in dynamic networks \cite{Frederic,Dalhia11}. In scientometrics, the knowledge maps have been generated from citation data to visualize the relationship between scholarly publications or disciplines. Early work on mapping journals focused on single disciplines. Morris \cite{Morris98} explored the interdisciplinary nature of medical informatics and its internal structure using inter-citation and co-citation analysis. Combination of the SCI and SSCI data was used in this study. McCain \cite{McCain98} performed the co-citation analysis for journals in neural network research. Cluster analysis, principal component analysis and multidimensional scaling (MDS) maps were used to identify the main research areas. Regarding to computer science, Ding \cite{Ding00} studied the relationship between journals in information retrieval area using the same techniques. Based on the ScieSearch database, Tsay \cite{Tsay03} mapped semiconductor literature using co-citation analysis. The datasets used in these studies were rather small, ranging from tens to several hundred journals. In more recent work, Boyack \cite{Boyack07} mapped the structure and evolution of chemistry research over a 30-year time frame. Based on a general map generated from the combined SCIE and SSCI from 2002, he assigned journals to clusters using inter-citation counts. Journals were assigned to the chemistry domains using JCR categories. Then, the maps of chemistry at different time periods and at domain level were generated. Maps show many changes that have taken place over the 30 years development of chemistry research.

Recently, several maps based on large-scale digital libraries have been published. ISI has published journal citation reports for many years. This dataset allows for generating the map of all of sciences. Leydesdorff has used the 2001 JCR dataset to map 5,748 journals from the SCI \cite{leydesdorff04} and 1,682 journals from the SSCI \cite{Leydesdorff042} in two separate studies. In those studies, Leydesdorff used Pearson correlation on citation counts as the edge weight and progressive lowering threshold to find the clusters. These clusters can be considered as disciplines or sub-disciplines. Moya-Aneg\'{o}n et al. \cite{Moya04} created category maps using documents with a Spanish address and ISI categories. The high level map shows the relative positions, sizes and relationships between 25 broad categories of science in Spain. Boyack \cite{Boyack05} combined SCIE and SSCI from 2000 and generated maps of 7,121 journals. The main objective of this study was to evaluate the accuracy of maps using eight different inter-citation and co-citation similarity measures.

There are several studies which applied SNA measures to derive useful information from knowledge maps. Leydesdorff \cite{Leydesdorff06} used the combination of SCIE and SSCI, and generated centrality measures (betweenness, closeness and degree centrality). These measures were analyzed in both global (the entire digital library) and local (small set of journals where citing is above a certain threshold) environments. Bollen et al. \cite{Bollen09} generated the maps of science based on clickstream data logged by six web portals (Thomson Scientific, Elsevier, JSTOR, Ingenta, University of Texas and California State University). They validated the structure of the maps by two SNA measures: betweenness centrality \cite{Wasserman95} and PageRank \cite{Brin98}. In another study, Bollen \cite{Bollen092} performed a principal component analysis on 39 scientific impact factors, including four SNA factors (degree centrality, closeness centrality, betweenness centrality and PageRank).

Regarding to the research on the performance of individuals and their local social network structures, Shi et al. \cite{Shi10} studied the citation projection graphs of publications in different disciplines, including natural science, social science and computer science, to understand their citation behaviors. Using several social network analysis measures, they identified the idiosyncratic citers, within-community citers and brokerage citers. They found that there are significant differences in how high, low and medium impact papers position their citation. There are also other studies on the optimal network structure for the individuals' performance \cite{Lambiotte09}, the benefits of the communities in fostering trust, facilitating the enforcement of social norm and common culture \cite{Coleman88}, and the benefits of structural holes and weak ties in accessing new information and ideas \cite{granovetter83}.

\section{The Role of Conferences in Computer Science Research}
Computer science history can be traced back from 1936, with the invention of Turing machine. Till early 1970s, the main publication outlet is journals. The Journal of Symbolic Logic (born in 1936), IEEE Transactions on Information Theory (1953), Journal of the ACM (1954), Information and Computation (1957) and Communications of the ACM (CACM) (1959) are probably the oldest journals in computer science. In late 1960s and early 1970s, some conferences emerged. IFIP Congress (1962), SYMSAC(Symposium on Symbolic and Algebraic Computation) (1966), the ACM Symposium on Operating Systems Principles (SOSP) (1966), Symposium on Operating Systems Principles (SOSP) (1967), International Joint Conference on Artificial Intelligence (IJCAI) (1969), Architecture of Computing Systems (ARCS) (1970), International Colloquium on Automata, Languages and Programming (ICALP) (1972), Symposium on Principles of Programming Languages (POPL) (1973) are some examples of the earliest conference series. Since early 1980s, conference has had a dominant present in computer science. According to DBLP digital library, as of 2010 there are 2716 conference series and 774 journals.

In 2009 and 2010, a dozen of articles, letters and blog entries discussed about the role of conferences, the quality and impact of conference publications \cite{Vardi09,Meyer09,Fortnow09}. In \cite{CACMStaff09}, Menczer supports the abolition of conference proceedings altogether and submissions should instead go to journals, which would receive more and more better ones. The impact and quality of conference publications are also questioned, mainly dues to the review process. Every conference has a desire to be ``competitive´´ and reducing the acceptance rate is an easy way. The great papers always are accepted and the worst papers mostly get rejected, but the problem here is for the vast majority of papers landing in the middle. That leads to an emphasis on safe papers (incremental and technical) versus those that explore new models and research directions outside the established core areas of the conferences \cite{Fortnow09}. Nevertheless, recent study by J. Chen and J. Kostan \cite{Chen10} shows that within ACM, papers in highly selective conferences are cited at a rate comparable to or greater than ACM transactions and journals. Freyne et al. \cite{Freyne10} demonstrates that papers in leading conferences match the impact of papers in mid-ranking journals and surpass the impact of papers in journals in the bottom half of the Thompson Reuters rankings.

Why conference becomes an important outlet in computer science? In \cite{Fortnow09}, L. Fortnow gave a short history of computer science conferences and the reasons for that computer science holds conferences. The fundamental reason is that the quick development of the field requires a rapid review and distribution of results. A complete journal publishing decision takes at least one year, comparing to 6 months for publishing in a conference. That delay is unacceptable for such a fast-changing field. Secondly, conferences bring the community together to disseminate new research and results, to network and discuss about the issues. That rarely happens in journals, where the only possible communication is between reviewers, editorial board and authors in review process. Lastly, with the tremendous continual growth in computer science, there are too many papers to publish and archival journals alone can not handle.

Our work is based on the above intuitions. We show that analysis on journal only can not fully capture the characteristics and development of computer science research since focusing exclusively on journal papers misses many significant papers published by conferences. We further show that conferences facilitate the communication and build a community between participants.
\section{Data Collection}
The dataset used in our study is the combination of DBLP and CiteSeerX digital libraries. We choose them because they cover most of sub-disciplines, while IEEE Xplore and ACM Portal cover only IEEE and ACM journals and conference proceedings. We retrieve the publication list of journals/conferences from DBLP. Unfortunately, DBLP does not record citations. Therefore, we use CiteSeerX to fill the citation list of publications in DBLP.

DBLP data was downloaded in July, 2009 which consists of 788,259 author's names, 1,226,412 publications and 3,490 venues. At the same time, we obtained CiteSeerX data by first download the OAI (Open Archives Initiative) dataset using the OAIHavester API. Since the OAI dataset contains only references between publications which are stored in CiteSeerX (with PDF documents), we continued to crawl XML documents from CiteSeerX site to obtain full citation list for each publication. Overall, we had complete CiteSeerX data with 7,385,652 publications (including publications in reference lists), 22,735,140 references and over 4 million author names.
\begin{figure}[h]
\centering
\includegraphics[height = 2.5in, width= 3.2in]{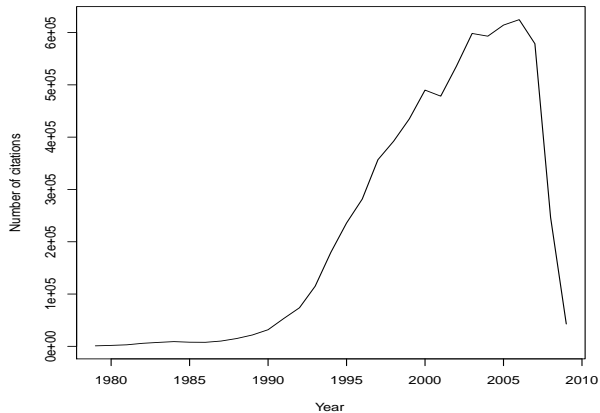}
\caption{Citation distribution}
\label{fig:no1}
\end{figure}

Naming is a problem that many digital libraries are faced because one author may have several names (synonyms) or there are several authors with the same name (homonyms). For example, in DBLP we can find seven authors with the name Chen Li. Consequently, several techniques have been developed for naming problem in digital libraries \cite{Han04,Pereira09,Lee05,Treeratpituk09}. In our analysis, we realize on the approaches that are implemented in CiteSeerX \cite{Huang06} and DBLP \cite{Ley09} and consider that the authors in these databases are identical.

We combined DBLP and CiteSeerX using a simple technique called \emph{canopy clustering} \cite{MacCallum00}. The basic idea of is to use a cheap comparison metric grouping records into overlapping clusters called \emph{canopies}. After that, records in the same cluster are compared using more expensive (and more accurate) similarity measures. We employed this idea to solve our problem. Firstly, publications in DBLP and CiteSeerX are clustered using the last name of authors. It can be argued as to whether the last name of authors give us the correct clusters, since one name can be expressed differently (e.g. Michael Ley vs. Ley Michael). However, in most cases author names of the same papers are presented in the same way in both digital libraries. In the second step, we used two similarity metrics to compare paper titles in each cluster: one less expensive \emph{Jaccard similarity} to filter out papers which are clearly un-matched, another more expensive \emph{Smith-Waterman distance} to correctly identify pair of matched papers. The process was implemented in Java using the SecondString\footnote{\url{http://secondstring.sourceforge.net/}} library and an Oracle database.

Overall, the matching algorithm gave us 864,097 pairs of matched publications, meaning about $70\%$ publications in DBLP were matched to publications in CiteSeerX. On average, each venue cites others 2306 times and is cited 2037 times. The distribution of the citations over years is given in Fig. \ref{fig:no1}, where the number of citations in 2009 and 2010 are low, simply because new publications are not crawled by CiteSeerX yet. It is not known whether this result reflects the real coverage of DBLP and CiteSeerX. However, in our experience lots of publications in CiteSeerX are not indexed in DBLP. The reason is that DBLP does not index some publication types such as pre-prints, in-prints, technical reports and letters, and it covers a limited number of PhD theses, master theses and books. That does not affect our analysis since we focuss on journal and conference publications. On the other hand, not all publications in DBLP are indexed by CiteSeerX. If a publication is not online and public, it will not be crawled by CiteSeerX.

\section{Networks Creation}
We created two networks using the dataset described above: one \emph{knowledge network K} based on relatedness of venues and one \emph{citation network F} based on citation counts. We processed as following:

 Bibliography coupling counts were calculated at the publication level on the whole digital libraries. These counts were aggregated at the venue level (3,490 venues), giving us the bibliography coupling counts between pairs of venues. Of 3,490 venues, 303 venues which have no citations were excluded. The result is a symmetric bibliography coupling frequency matrix $V$ with venues as columns and rows. Based on this matrix, we created the \emph{knowledge network} $K$ by normalizing bibliography coupling counts using cosine similarity as suggested in \cite{Klavans06}, in which the full version of cosine index was used. Concretely, cosine similarity between pair of venues is computed as:
\begin{equation}
C_{i,j} = \frac{\overrightarrow{B_{i}}\bullet\overrightarrow{B_{j}}}{\parallel\overrightarrow{B_{i}}\parallel\times\parallel\overrightarrow{B_j}\parallel}
= \frac{\displaystyle\sum_{k=1}^{n}B_{i,k}B_{j,k}}{\sqrt{\displaystyle\sum_{k=1}^{n}B_{i,k}^2}\sqrt{\displaystyle\sum_{k=1}^{n}B_{j,k}^2}}\
\end{equation}
where $C_{i,j}$ is the cosine similarity between venue $V_i$ and $V_j$, $B_i$ is the vector representation of the list  of citations from venue $V_i$ to all publications, $n$ is number of publications in the database, and $B_{i,k}$ is the number of times venue $V_i$ cites publication $k$. The resulting network consists of 1,930,471 un-directed weighted edges. 120 venues whose cosine similarity to others equal to zero were not included in the network.

The \emph{citation network} $F$ is formulated by counting the inter-citation between venues. Nodes are venues and there is an edge from venue $V_i$ to venue $V_j$ if $V_i$ cited $V_j$, weighted by number of that citations. The network contains 351,756 directed edges, resulting in a network density of 3.5\%.

To prevent noise in the visualization and analysis, we consider the most relevant connections between venues. For the knowledge network $K$, we eliminated all connections which have cosine similarity smaller than $0.1$, obtaining the reduced network $K'$ whose connection cosine similarity is in the range [$0.1$, $1.0$]. Although this threshold is arbitrary, the network $K'$ retains 1,739 nodes and 9,637 connections, corresponding to 57\% of the nodes and 0.5\% of the edges of the original network. For the citation network $F$, the same procedure was performed in which we only keep the connections whose citation counts were greater than 50. The remaining network $F'$ contains 1,060 venues and 9,964 connections, corresponding to 33\% of the nodes and 2.8\% of the edges of the original network. A summary of networks properties is given in Table \ref{tab1}.
\begin{table}[tbh]
\centering
 \caption{\label{tab1}Networks Summary}
\begin{tabular}{|p{2.3cm}p{1cm}p{0.8cm}p{1.3cm}p{1cm}|}
    \hline
    \textbf{Property} & \textbf{F} & \textbf{F'} & \textbf{K}&\textbf{K'}\\
    \hline
    Nodes&3,187&1,060 &3,067&1,739\\

    Edges&351,756&9,964&1,930,471&9,637\\
    Components&1&6&1&71\\
    Density&3.5\%&0.89\%&20\%&0.3\%\\
    Clustering coef.&0.569&0.764&0.786&0.629\\
    \hline
\end{tabular}
\end{table}

The reason for creating two networks is as follows. Because of the diversity of publication types and interdisciplinary nature of computer science, publications often refer to the publications (e.g. preprints, letters) which may not be published by any journals, conferences or workshops. The references also point to the publications in other disciplines. For example, lots of papers on SNA cite the work done by Newman and Barabasi which are published in science journals (Phy. Rev. Letters or Nature). That should be considered when calculating the similarity between venues. Therefore, we computed the cosine similarity on the complete list of references at the paper level, then aggregated at the venue level to create the knowledge network. However, to study the information diffusion and the impact of venues in the domain, we need only the citation counts between themselves. The citation network was created based on the inter-citation counts between venues, accordingly.
\section{Knowledge Network Visualization}
\begin{figure*}
\centering
\fbox{
\includegraphics[height = 4.7in, width= 4.7in]{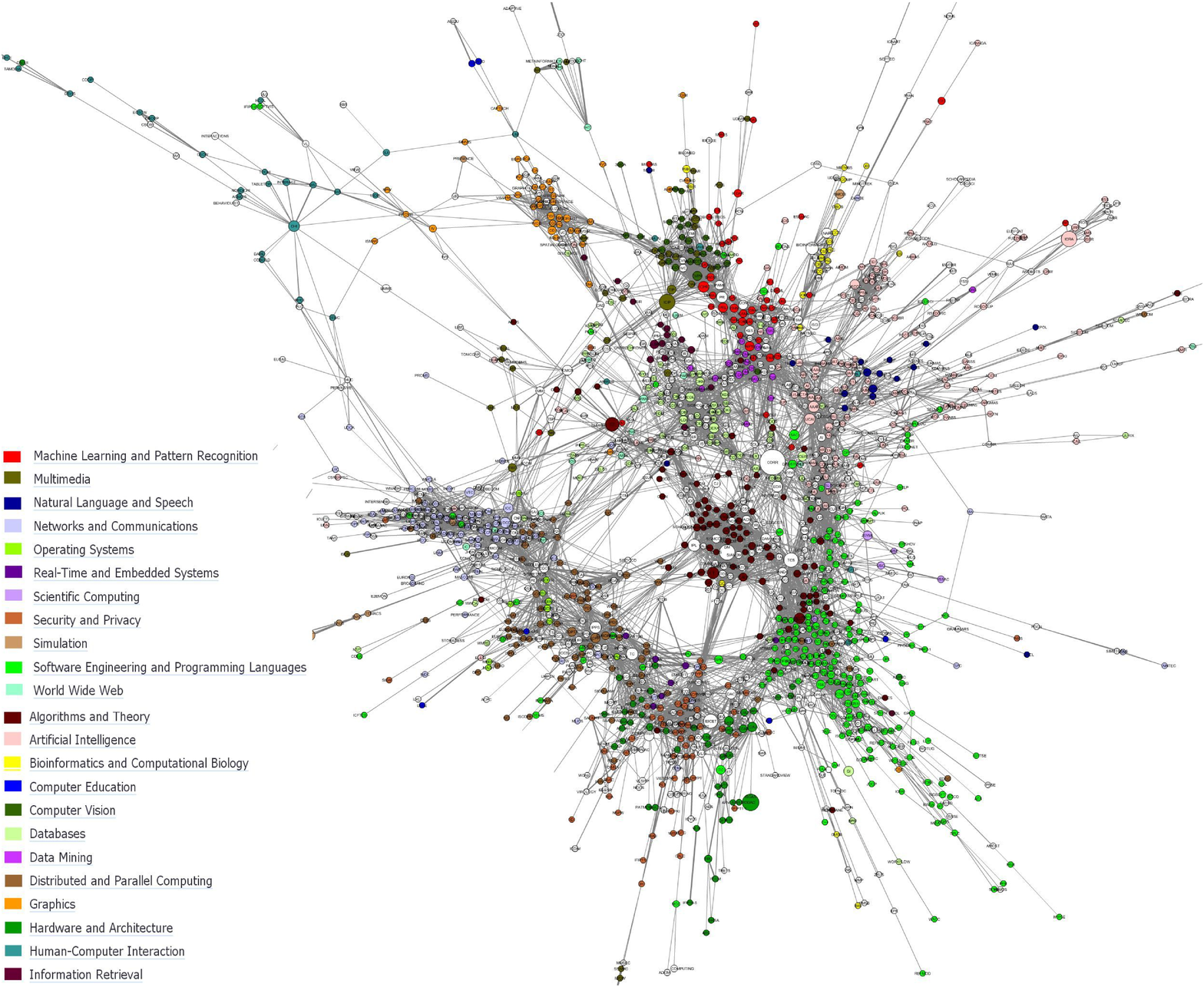}
}
\caption{The combined knowledge network (giant component)}
\label{fig:no2}
\end{figure*}

\begin{figure*}
\centering
\fbox{
\includegraphics[height = 4.7in, width= 4.7in]{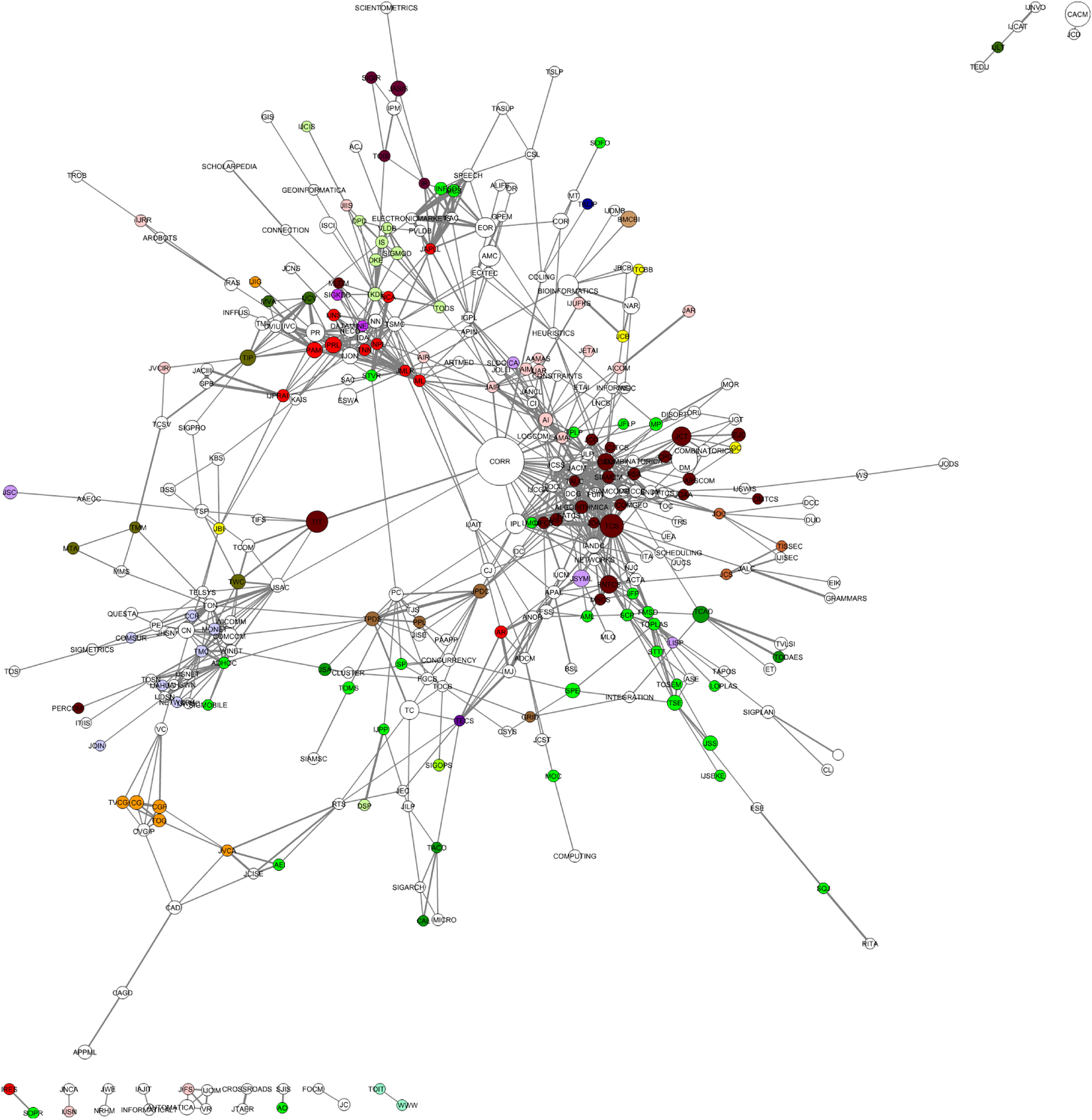}
}
\caption{The knowledge network using journals only}
\label{fig:no3}
\end{figure*}
We visualized the knowledge network $K'$ using \emph{smart organic layout} implemented in the yFiles\footnote{\url{http://www.yworks.com/}} library, based on the force-directed paradigm \cite{fruchterman91}. The visualization is given in Fig. \ref{fig:no2}, where venues are represented as circles with diameter denoting the number of publications and the thickness of connections denotes the cosine similarity. Nodes are colored according to their assignment to domain categories in Microsoft Academic Search\footnote{\url{http://academic.research.microsoft.com/}}(Libra). White color nodes are un-categorized venues. Libra assigns 2637 venues to 23 domains, so 430 venues in our database remain un-categorized. We also accounted that some venues are assigned to multiple domains. For those, we randomly chose one of the assigned domains. Fig. \ref{fig:no3} gives the visualization of the knowledge network using journals only, which allow us to compare the visual structures of the two networks.

Any interpretation of the visual structure of the knowledge network in Fig. \ref{fig:no2} has to take into account the following considerations. Firstly, different iterations of force-directed algorithm can converge on different visualizations of the knowledge network. Fig. \ref{fig:no2} is not the only or best possible visualization. It is selected because it represents a clear visualization of connections between venues in the knowledge network and its main structural features were stable across many iterations of the visualization algorithm. Secondly, the force-directed algorithm groups together venues that are strongly connected in the knowledge network. The appearance of clusters is thus depends on the weight of the connections in the knowledge network and is not the artifact of the visualization. Finally, the exact geometric coordinates of journals/conferences and clusters vary depending on the visualization algorithm and are thus considered artifacts of the visualization.

Fig. \ref{fig:no2} shows us a clear cluster structure in which venues in the same domain are placed in clusters. In contrast, the network of journals only (Fig. \ref{fig:no3}) is little ``un-ordered`` and one can not identify sub-disciplines from this network. In Fig. \ref{fig:no2}, large and coherent clusters are algorithms and theory, artificial intelligence, software engineering, security and privacy, distributed and parallel computing, networks and communications, computer graphics, computer vision, databases, data mining and machine learning. They cover most of the core topics of computer science. Some domains do not have their own clusters. Venues in those domains are placed in the same clusters with venues from closely related domains. For example, data mining and machine learning are combined in one cluster; information retrieval sticks to databases; natural language and speech processing is a sub-group of the artificial intelligence cluster etc. That result reflects the hierarchical structure of domain classification.

Connections between venues in the network cross multiple domains. Dominating in the middle of the network are venues in algorithms and theory. This domain are connected to many other domains in the border of the wheel. The second dominator at the center is databases. In clockwise order, starting at 12AM, databases is tightly connected to information retrieval, data mining and machine learning (1PM), artificial intelligence (2PM), as well as software engineering (the green color, at 3PM to 4 PM). Computer graphics connects to computer vision, multimedia and human-computer interaction studies. We can also easily identify the cluster of bioinformatics which has connections to artificial intelligence, data mining and machine learning. At the bottom of the wheel, there is a mixed cluster of venues from hardware and architecture, real-time and embedded systems, security and privacy. This cluster connects strongly to software engineering and distributed computing.

\begin{figure*}
\centering
\fbox{
\includegraphics[height = 3.7in, width= 4.5in]{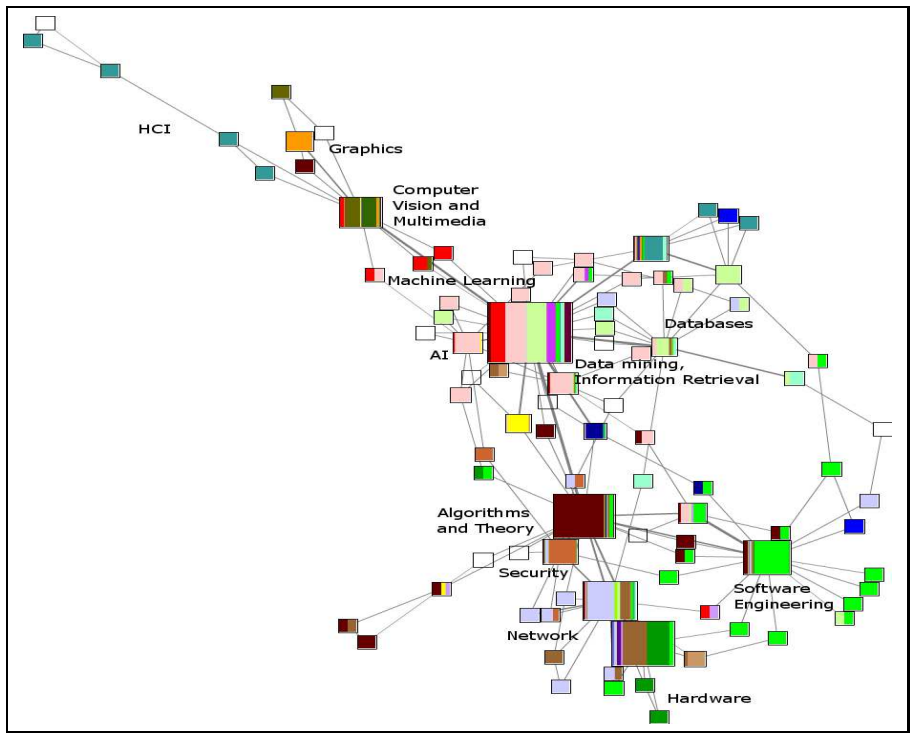}
}
\caption{The knowledge network at a cluster level}
\label{fig:no4}
\end{figure*}

Although the visualization of the knowledge network at venues level shows us a clear cluster structure, it would be more pleasant to see the visualization at the cluster level. During the network reduction process, lots of venues were excluded. To make the visualization at cluster level more precise, we process as follows:
 \begin{itemize}
   \item The knowledge network $K'$ is clustered using a density-based clustering algorithm proposed by Newman and Clauset \cite{newman-2004-692,clauset-2004-70}. The basic idea of the algorithm is to find a division of the network into clusters within which the network connections are dense, but between which they are sparser. To measure the quality of a division, the modularity $Q$ \cite{newman-2006-103,newman-2004-69} is used. In our case, the algorithm gives us 92 clusters with the modularity $Q=0.771$.
   \item  Using the bibliography coupling frequency matrix $V$ where columns and rows are venues, the counts were aggregated to cluster level for the venues which were assigned to clusters, thus give us the bibliography coupling counts between un-clustered venues and clusters. That results in a bibliography coupling frequency matrix $V'$ with venues and clusters in both columns and rows. We calculate the cosine index between 1328 un-clustered venues and 92 clusters, and assign un-clustered venues to clusters with which they have highest cosine values.
   \item After that, cosine index is re-computed for pairs of clusters in the same way as we did for venues.
 \end{itemize}
Fig. \ref{fig:no4} the visualization at cluster level where clusters are squares with the size denoting the number of venues and the weight of the connection between clusters is the cosine similarity. Clusters are colored using the same color scheme as in Fig. \ref{fig:no2}. The colors show the fraction of domain venues in clusters. To prevent clutter, for each cluster we retain the 2 strongest outbound relationships. The network is manually labeled based on the assignment of clusters to particular domains.

The network in Fig. \ref{fig:no4} can be interpreted as follows. In general, the appearance of the network is similar to the network in Fig. \ref{fig:no2}. Most of domains are assigned to more than one clusters in which they dominate or share the ``power`` with other related fields. The exceptions are graphics and bioinformatics which are uniquely assigned to one cluster. Large clusters are composed of several closely related domains (except for the large clusters of algorithms and theory, and software engineering, where the venues of these fields dominate the clusters). For example, one cluster in the upper half of the diagram contains machine learning, AI, databases, data mining, information retrieval and the world wide web. These fields seem to be very exciting research areas with one large cluster and many small ones closely connected to each other. AI is the most interdisciplinary area. Venues in this field are distributed in multiple clusters which have many connections to other areas such as databases, data mining, information retrieval, machine learning, WWW, software engineering, algorithms and theory, bioinformatics and HCI. Computer vision, multimedia and graphics are quite marginal topics which have relationships only to machine learning.
\section{The Evolution of the Knowledge Network}
The visualization given in Fig. \ref{fig:no2} is useful for observing the recent organization of the computer science knowledge. Now the interesting question is that how the computer science knowledge comes to this stage. In particular, we would like to see the development of the research areas - how the new fields emerge and develop over time, how venues come together to form clusters (sub-disciplines) and how they split into sub-clusters, how new venues are connected to the existing venues and clusters, and how the strength of the connections between clusters increases to form the "shape" of the computer science knowledge.

To answer these questions, we visualize the knowledge network at different time points, from 1990 to 2005, with 5-year intervals using the same technique as presented in Section 3 and Section 4. To compute the similarity between venues at a certain time, we consider only the papers published from this time point backwards. The scales for cosine similarities (the thickness of connections) and venue size (node size) have been kept constant to enable easy inspection of the changes. Note that the assignment of venues to sub-disciplines by Libra is not perfect, so there are some misclassifications. To interpret the visualization, we have to base on both the clusters of venues grouped by the visualization algorithm and the sub-discipline labels, where in each cluster, if a sub-discipline has a dominant number of venues then this cluster represents that sub-discipline.

\begin{figure*}
\centering
\fbox{
\includegraphics[height = 5.5in, width= 4.7in]{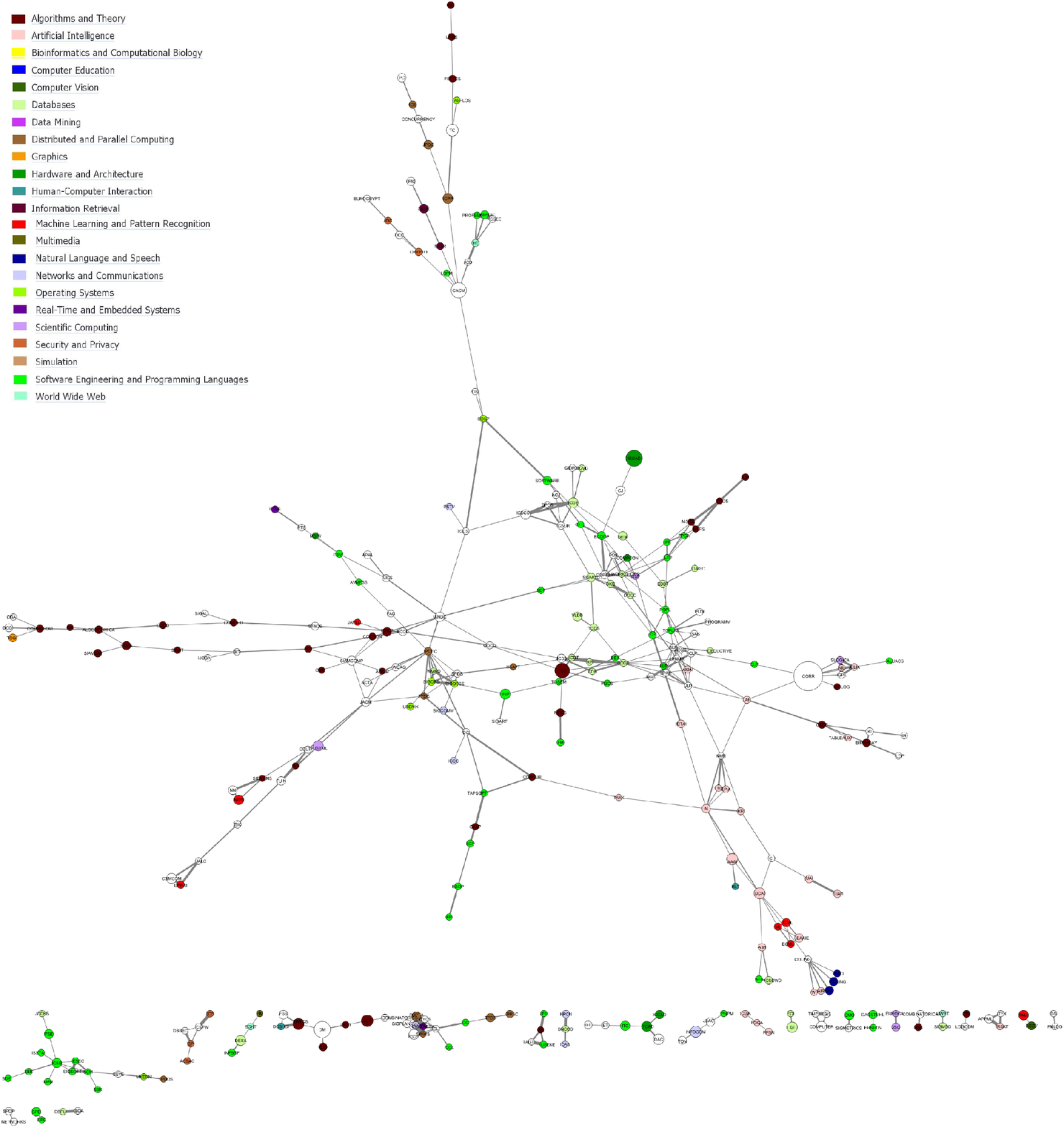}
}
\caption{The knowledge network in 1990}
\label{fig:no5}
\end{figure*}
\begin{figure*}
\centering
\fbox{
\includegraphics[height = 5.5in, width= 4.7in]{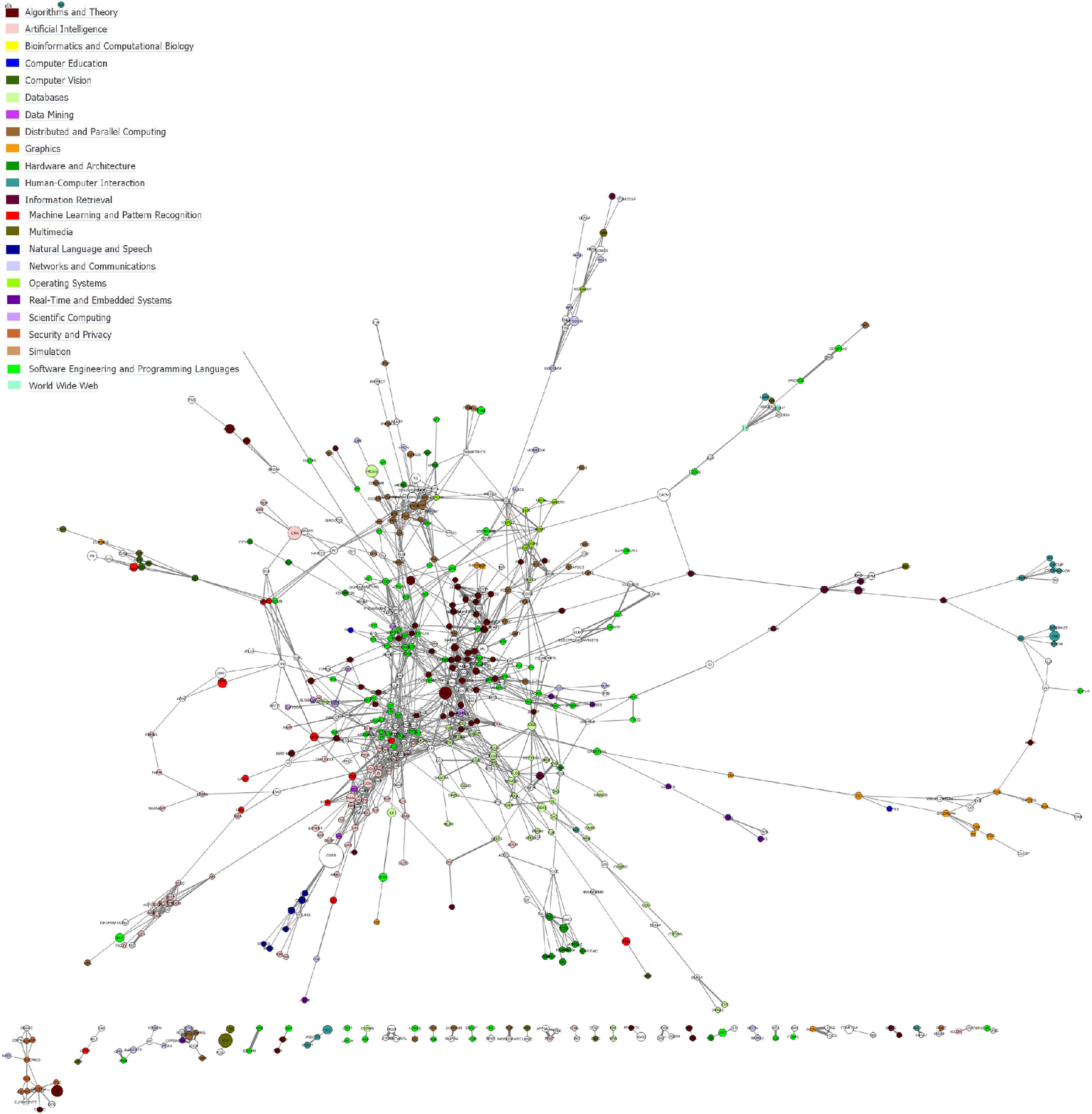}
}
\caption{The knowledge network in 1995}
\label{fig:no6}
\end{figure*}
\begin{figure*}
\centering
\fbox{
\includegraphics[height = 5.5in, width= 4.7in]{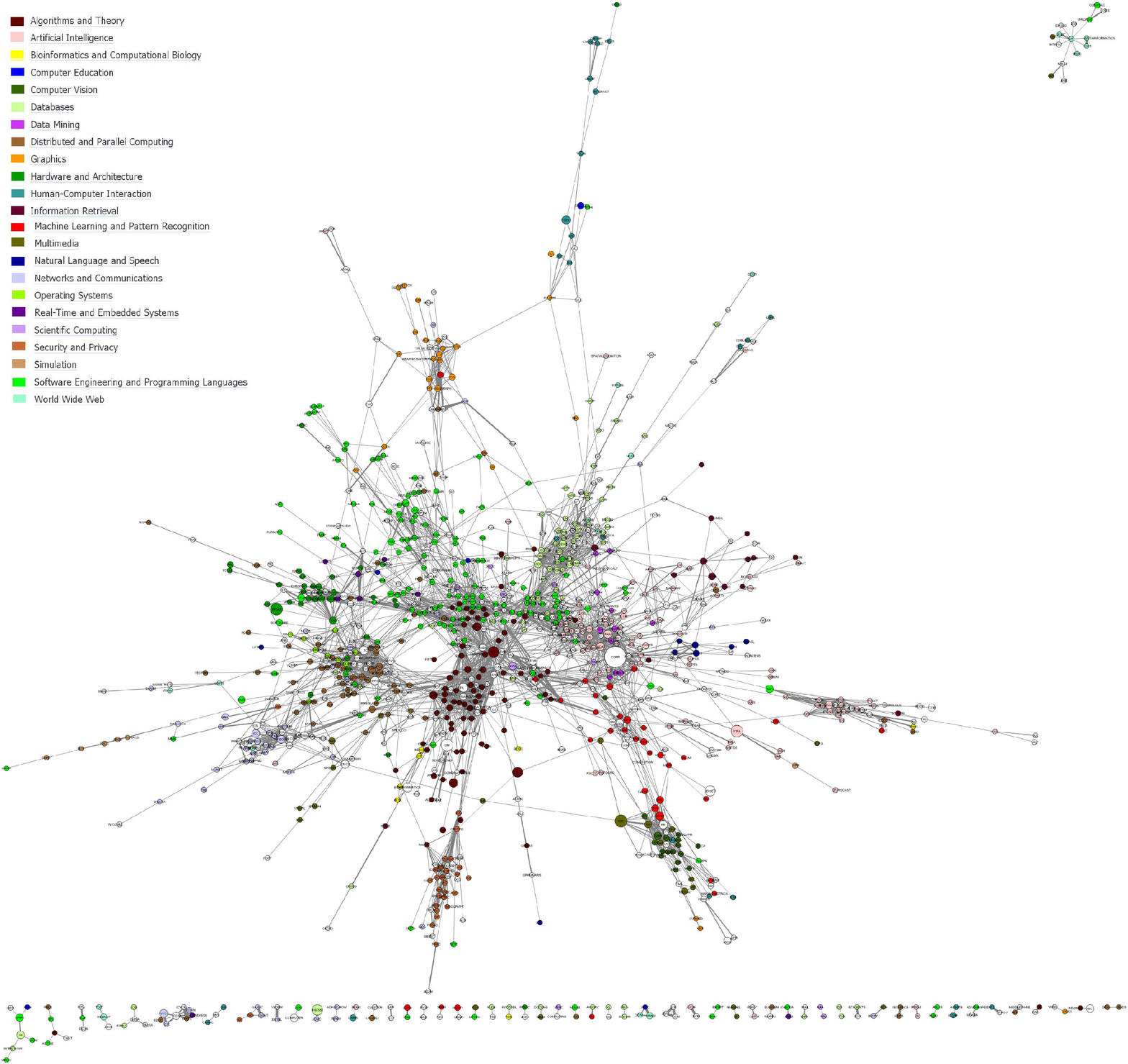}
}
\caption{The knowledge network in 2000}
\label{fig:no7}
\end{figure*}
\begin{figure*}
\centering
\fbox{
\includegraphics[height = 5.5in, width= 4.7in]{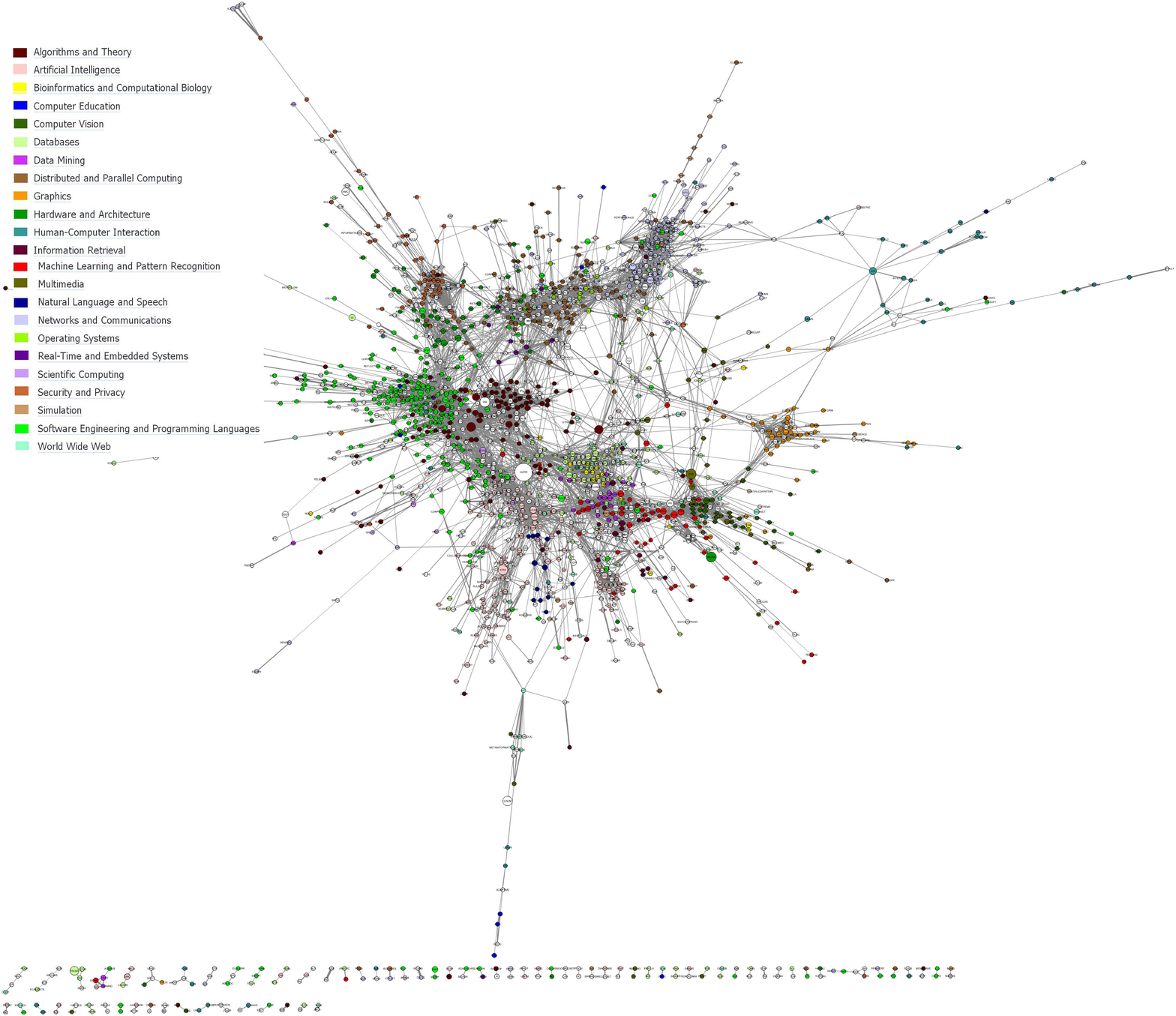}
}
\caption{The knowledge network in 2005}
\label{fig:no8}
\end{figure*}

The visualizations of the knowledge network in 1990, 1995, 2000 and 2005 are given in Fig. \ref{fig:no5}, \ref{fig:no6}, \ref{fig:no7} and \ref{fig:no8}. A close inspection of these figures and Fig. \ref{fig:no2} reveals many changes. In 1990, the knowledge network is not clearly clustered. Although we can identify the groups of venues in some sub-disciplines such as database, artificial intelligent, algorithms and theory, software engineering and programming languages, privacy and security, the venues in these domains are distributed in several groups and the connection in these groups is very sparse (low density). Some sub-disciplines even are separated into disconnected components (e.g. a group of software engineering venues at the bottom left corner). In 1995 there are still some disconnected groups, but venues start to come closer to form the core of sub-disciplines. We can also observe the early connections between fields. At the center, there is a large body of algorithms and theory which has many connections to other large clusters such as software and programming languages, database, artificial intelligence, distributed and parallel computing. Computer graphics (on the right hand) starts a cluster and has connections to human-computer interaction. Some other sub-disciplines emerge, such as machine learning and data mining emerge from artificial intelligence, networking separates from operating system and distributed and parallel computing. Well established venues (shown in Fig. \ref{fig:no5}) continue to play the central role in the domains, such as VLDB, SIGMOD and TKDE in database, TSE and ICSE in software and programming languages, SIGCOMM and INFORCOM in networking, AAAI, AI and IJCAI in artificial intelligence.

We observe these trends also in 2000 and 2005 where sub-disciplines become more organized and mature. The connections between sub-disciplines are also become clearer, reflecting the interdisciplinary nature of computer science research. Sub-disciplines are also starting to separate, as we see in Fig. \ref{fig:no7} and Fig. \ref{fig:no8} where artificial intelligence is divided into several clusters, or bioinformatics emerge from database research. However, merging seems to dominate the development trend, where disconnected components of the network join to the giant component. For example, in 1990 there are several disconnected components of software engineering, privacy and security as well as other domains. In 1995, a big disconnected component of software engineering joins the giant component. The component of privacy and security stays disconnected, but becomes bigger. Then in 2000, that component finally joins the giant component. It is interesting to connect these observations to what actually happened at that time. For example, in 1996, HTML 2.0 specification was maintained as a standard and in 1997, it became an international standard (RFC 2070). That is the reason for the blow of the Internet with many commercial software vendors and platforms, especially Internet Explorer developed in the Windows 95 system. Before that time, privacy and security was quite an isolated research domain in computer science. However, with the increasing use of the Internet where people can exchange information quickly and freely, security becomes one of the main concerns and attracts a lot of attention from both industry and research community. That could be the reason why in 1995, security and privacy research stays as a disconnected component, but in 2000 it connects to the giant component and is one of the large clusters.

The visualizations in Fig. \ref{fig:no5}, \ref{fig:no6}, \ref{fig:no7}, \ref{fig:no8} reveal a lot of information, more than we can describe here. Thus, we highlight the main changes and trends. Over time, main topics in computer science, including algorithms and theory, artificial intelligent, database, networking, and software engineering, develop consistently. Domains become more and more interdisciplinary where they connect more or less to other domains or sub-domains. Fields are starting to split into sub-fields, though merging dominates the development trend. New fields or sub-fields continuously emerge from the existing sub-disciplines. With the growth of the Web, data mining and information retrieval, emerging from database and artificial intelligence, as well as privacy and security are becoming more and more exciting fields with a lot of conferences and journals.
\section{Venue Ranking}
There are different metrics to evaluate the performance and prestige of individuals and journals such as citations count, H-index \cite{Hirsch05}, impact factor\footnote{\url{http://thomsonreuters.com/products_services/science/academic/impact_factor/}} by the Institute for Scientific Information (ISI), now part of Thomson Reuters. However, these metrics are controversial \cite {Seglen97,Kumar09} and it turns out that using one single metric cannot judiciously evaluate the impact of a journal or a scientist. A metric should be used in the combination with other metrics to fully and reliably justify the performance of scientists, publications and venues.

We employ two social network measures, \emph{betweenness centrality} and \emph{PageRank}, for venue ranking. These measures do not intend to be a replacement, but a complement to the existing metrics. Given the assignment of venues to domains and the citation network $F'$, we calculated \emph{node's betweenness centrality} and \emph{PageRank} to determine interdisciplinary and high prestige venues, respectively. Using PageRank has one advantage over the impact factor: PageRank does highly rank venues which are cited by other highly ranked venues, so new venues have a higher impact when they are cited by well-known venues.
%In social network analysis, there are different centrality measures such as betweenness, closeness and degree centrality.

Betweenness centrality \cite{Wasserman95} of a venue $V_i$ is defined as the number of shortest paths in the network that pass through $V_i$ and it is computed as follows:
\begin{equation}
C(V_k) = \displaystyle\sum_{i\neq j\neq k}\frac{P_{i,j}(k)}{P_{i,j}}\
\end{equation}
where $P_{i,j}$ is the number of (weighted) shortest paths between venues $V_i$ and $V_j$, $P_{i,j}(k)$ is the number of that shortest paths which go through venue $V_k$. Highly value of betweenness centrality indicates a venue as a "gateway" which connects a large number of venues and venue clusters. Venues with high betweenness centrality values often are interdisciplinary. Table \ref{tab2} gives the list of top 30 centrality venues. They are indeed highly interdisciplinary venues. The first position is CORR (Computing Research Repository) with the betweenness 0.185. DBLP classifies it as a journal, but in fact CORR is a repository to which researchers could submit technical reports. CORR covers almost every topic of computer science. Papers published in CORR are not peer reviewed, only the relatedness to the topic area is checked. That is the reason for the appearance of CORR as a large venue in the visualization (Fig. \ref{fig:no2}) and as a top interdisciplinary venue. Among others, AI, machine learning, databases and the world wide web contribute ten venues to this list. That confirms their interdisciplinary nature reflected in Fig. \ref{fig:no4}.

\begin{table*}[tbh]
\centering
 \caption{\label{tab2}Top betweenness centrality venues}
\begin{tabular}{|p{0.5cm}p{2cm}cp{4.5cm}|}
    \hline
    \textbf{Rank} &\textbf{Name} & \textbf{Type} & \textbf{Libra classification} \\
    \hline
1&CORR	&\emph{J}	&\emph{Un-categorized}\\
2&TCS	&\emph{J}	&Algorithms and Theory\\
3&INFOCOM	&\emph{C}	&Networks\&Communications \\
4&AI	&\emph{J}	&Artificial Intelligence\\
5&CSUR	&\emph{J}	&\emph{Un-categorized}\\
6&TC	&\emph{J}	&\emph{Un-categorized}\\
7&TSE	&\emph{J}	&Software Engineering \\
8&JACM	&\emph{J}	&\emph{Un-categorized}\\
9&CACM	&\emph{J}	&\emph{Un-categorized}\\
10&CHI	&\emph{C}	&Human-Computer Interaction\\
11&ML	&\emph{J}	&Machine Learning\\
12&IJCAI	&\emph{C}	&Artificial Intelligence \\
13&TOPLAS	&\emph{J}	&Software Engineering \\
14&AAAI	&\emph{C}	&Artificial Intelligence \\
15&PAMI	&\emph{J}	&\emph{Un-categorized}\\
16&ICRA	&\emph{C}	&Artificial Intelligence\\
17&SIAMCOMP	&\emph{J}	&\emph{Un-categorized}\\
18&TPDS	&\emph{J}	&Distributed\&Parallel Computing\\
19&ICDE	&\emph{C}	&Databases\\
20&WWW	&\emph{C}	&World Wide Web\\
21&TKDE	&\emph{J}	&Databases\\
22&CVPR	&\emph{C}	&Computer Vision\\
23&ENTCS	&\emph{J}	&Algorithms and Theory\\
24&VLDB	&\emph{C}	&Databases\\
25&IPPS	&\emph{C}	&Scientific Computing\\
26&ALGORITHMICA	&\emph{J}	&Algorithms and Theory\\
27&ICDCS	&\emph{C}	&Networks\&Communications\\
28&CAV	&\emph{C}	&Software Engineering \\
29&SIGGRAPH	&\emph{C}	&Graphics\\
30&CN	&\emph{J}	&Networks\&Communications\\
    \hline
\end{tabular}
\end{table*}

The PageRank score of a venue is computed according to the PageRank algorithm\cite{Brin98}. The algorithm iteratively calculates the PageRank score of a venue based on the score of its predecessors in the network as in the following equation.
\begin{equation}
P(V_i) = (1-d) + d \displaystyle\sum_{j}\frac{P(V_j)}{O(V_j)}\
\end{equation}
where $P(V_i)$ is the PageRank score of venue $V_i$, $V_j$ is the predecessor of $V_i$ and $O(V_j)$ is out-degree of $V_j$. Parameter $d$ is the \emph{dumping factor} which usually is set to 0.85 in literature. We note that the \emph{dumping factor} $d$ models the random Web surfer. Web surfing behavior is different to citing behavior, so the value of $d$ maybe different in our case. We use here the same value of $d$ and keep this note in mind.

The list of 30 highest PageRank venues is given in Table \ref{tab3} where column \emph{Type} denotes type of venue (\emph{J} for journal and \emph{C} for conference/workshop). PageRank favors venues that are well-connected to other well-connected venues. Surprisingly, CORR is in sixteenth position though it mostly consists of technical reports. The list in Table \ref{tab3} contains not only journals, but also the leading conferences in the fields. From the list, one can see the well-known venues such as Communication of the ACM (CACM), Journal of the ACM (JACM), Journal of Artificial Intelligence (AI), SIAM Journal on Computing (SIAMCOMP) and ACM Transaction on Computer Systems (TCS) as well as conferences in different fields such as SIGGRAPH, AAAI, SOSP, SIGCOMM, POPL, VLDB, NIPS etc. %The number of conferences in the list confirms their role in computer science.

\begin{table*}[tbh]
\centering
 \caption{\label{tab3}Top PageRank venues}
\begin{tabular}{|p{0.5cm}p{2cm}cp{4.15cm}|}
    \hline
    \textbf{Rank} &\textbf{Name} & \textbf{Type} & \textbf{Libra classification} \\
    \hline
1&CACM	&\emph{J}	& \emph{Un-categorized}\\
2&JACM	&\emph{J}	&\emph{Un-categorized}\\
3&AI	&\emph{J}	&Artificial Intelligence\\
4&SIAMCOMP	&\emph{J}	&\emph{Un-categorized}\\
5&TCS	&\emph{J}	&Algorithms and Theory\\
6&SIGGRAPH	&\emph{C}	&Graphics\\
7&TSE	&\emph{J}	&Software Engineering \\
8&JCSS	&\emph{J}	&\emph{Un-categorized}\\
9&AAAI	&\emph{C}	&Artificial Intelligence\\
10&SOSP	&\emph{C}	&Operating Systems\\
11&SIGCOMM	&\emph{C}	&Networks\&Communications\\
12&PAMI	&\emph{J}	&Machine Learning\\
13&INFOCOM	&\emph{C}	&Networks\&Communications\\
14&IJCAI	&\emph{C}	&Artificial Intelligence\\
15&POPL	&\emph{C}	&Software Engineering \\
16&CORR	&\emph{J}	&\emph{Un-categorized}\\
17&IANDC	&\emph{J}	&\emph{Un-categorized}\\
18&TOCS	&\emph{J}	&\emph{Un-categorized}\\
19&ISCA	&\emph{C}	&Hardware and Architecture\\
20&TC	&\emph{J}	&\emph{Un-categorized}\\
21&STOC	&\emph{C}	&\emph{Un-categorized}\\
22&VLDB	&\emph{C}	&Databases\\
23&ML	&\emph{J}	&Machine Learning\\
24&PLDI	&\emph{C}	&Software Engineering\\
25&TOPLAS	&\emph{J}	&Software Engineering\\
26&TON	&\emph{J}	&\emph{Un-categorized}\\
27&SODA	&\emph{C}	&Algorithms and Theory\\
28&NIPS	&\emph{C}	&Machine Learning\\
29&COMPUTER	&\emph{J}	&\emph{Un-categorized}\\
30&TIT	&\emph{J}	&Algorithms and Theory\\
\hline
\end{tabular}
\end{table*}

\section{Understanding the Collaboration and Citation Behavior}
\subsection{Venues Subgraphs}
To understand the collaboration and citation behavior of the communities of venues, we study the properties of the co-authorship and citation subgraphs of venues. We take all the papers published in a venue and extract its co-authorship network. The resulting network consits only the collaborations of the authors in the venue. Note that two authors might collaborate with each other in other venues, but might not collaborate in the venue under consideration. However, since we investigate the collaborations of authors working on the topics of the venue and how the venue maintains and cultivates these collaborations, it is not necessary to consider the collaborations of these authors in other venues. To create citation subgraphs, we take all the publications cited by papers published in a given venue, project them on the underlying citation graph and extract the subgraph of citations among these publications. Formally, we define the co-authorship subgraph $G_a = (A,E)$ of a venue is a graph where $A$ is the set of authors who published some papers in this venue and there is a connection $e \in E$ between author $a_i$ and $a_j \in A$ if they wrote a paper published in this venue together. Similarly, we define the citation subgraph of a venue $G_c = (P, C)$, where $P$ is the set of publications cited by papers published in this venue and $C$ is the set of citations among these publications.

\begin{figure}
\centering
\subfloat[Part 1][Network Type 1]{\fbox{\includegraphics[height = 1.5in, width= 2.3in]{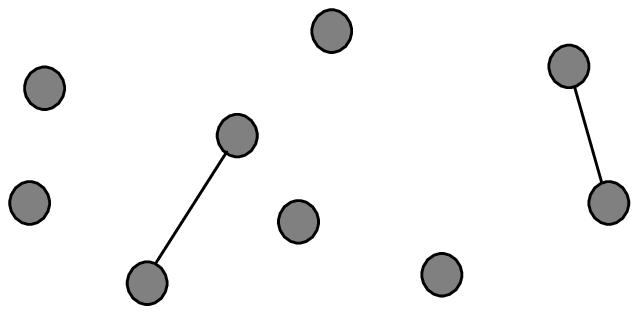}}
\label{fig:fig9-a}}
\subfloat[Part 2][Network Type 2]{\fbox{\includegraphics[height = 1.5in, width= 2.3in]{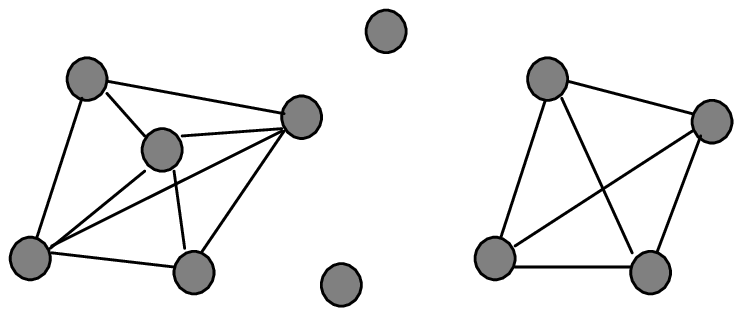}}
\label{fig:fig9-b}}\\
\subfloat[Part 3][Network Type 3]{\fbox{\includegraphics[height = 1.5in, width= 2.3in]{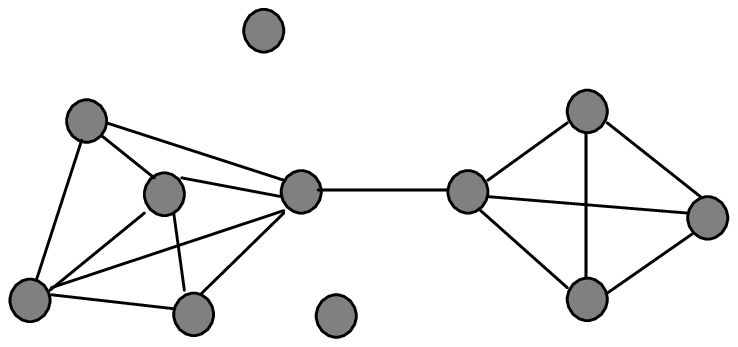}}
\label{fig:fig9-c}}
\subfloat[Part 4][Network Type 4]{\fbox{\includegraphics[height = 1.5in, width= 2.3in]{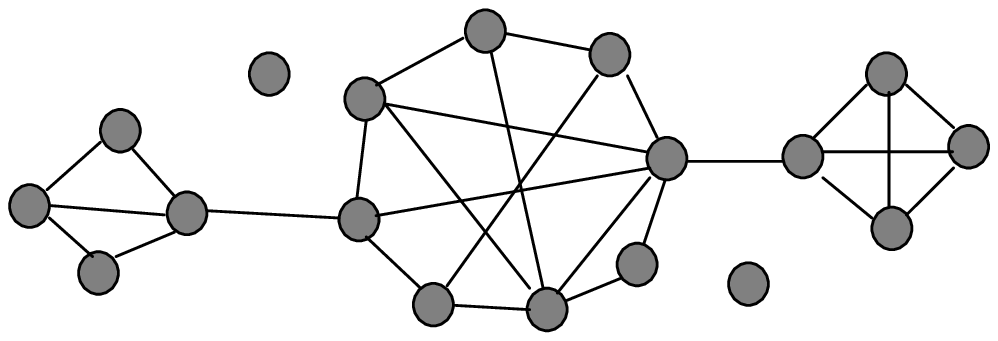}}
\label{fig:fig9-d}}\\
\caption{Network types}
\label{fig:no9}
\end{figure}

Given the co-authorship and citation subgraphs of venues, we then elaborate a set of network metrics that characterize and describe their structure. To give an idea about what type of networks we are trying to classify, let us take a look at the example given in Fig. \ref{fig:no9}. Fig. 9a (type 1) shows a network that is sparsely connected. The density of this network is rather low. In Fig. 9b (type 2), nodes are clustered in small disconnected components. Fig. 9c (type 3) describes a network where several small disconnected components come together to form a large connected component. In Fig. 9d (type 4), there exists a dense, large component and several small components connected to it. Intuitively, for citation subgraphs, network type 2 demonstrates the venues where the citations are placed in un-related sub-disciplines, meaning that it is unfocused. In network type 3, different clusters of papers that correspond to different sub-disciplines are cited but the connections between these sub-disciplines are also identified. Network type 4 illustrates a focused and interdisciplinary venue where the cited papers are clustered in a big largest component that can be considered as the main theme of the venue, and this component is connected to many other smaller components corresponding to the related sub-disciplines.

We employ four network metrics \cite{Wasserman95} in order to distinguish the four types of network. For every venue, we use these four metrics to characterize the features of its citation and co-authorship subgraphs. The four metrics are defined as follows:
\begin{itemize}
  \item \textbf{Density (M1)}: Density of a graph $G=(V,E)$ where $V$ is the set of vertices, $E$ is the set of edges, is defined as:
  \begin{equation}
    D(G) = \frac{2\mid E \mid}{\mid V\mid(\mid V\mid-1)}\
  \end{equation}

  \item \textbf{Clustering coefficient(M2)}: Local clustering coefficient of a node $v_i$ is defined as follows:
   \begin{equation}
    C(v_i) = \frac{\mbox{number of closed triads connected to }v_i}{\mbox{number of triples of vertices centered on }v_i}\
  \end{equation}
  The average local clustering coefficient is defined as
  \begin{equation}
    C = \frac{\sum C(v_i)}{\mid V\mid}\
  \end{equation}

  \item \textbf{Maximum betweenness (M3)}: is the highest betweenness of the nodes in $G$. The betweenness of a node $v_i$ is defined as
  \begin{equation}
    B(v_i) = \sum_{j\neq k} \frac{\sigma^{v_i}(v_j,v_k)}{\sigma(v_j,v_k)}\
  \end{equation}
  where $\sigma^{v_i}(v_j,v_k)$ is the number of shortest paths from node $v_j$ to node $v_k$ that pass through $v_i$ and $\sigma(v_j,v_k)$ is the total number of shortest paths from $v_j$ to $v_k$. The betweenness may be normalized by dividing through the number of pairs of vertices not including $v_i$, which is $(n - 1)(n - 2)$ for directed graphs and $(n - 1)(n - 2)/2$ for undirected graphs, where $n$ is the number of vertices in the network.

  \item \textbf{Largest connected component (M4)}: the fraction of nodes in the largest connected component.
\end{itemize}
To summarize, the four metrics allow us to differentiate the four types of network based on the scheme in table \ref{tab4}.
\begin{table}[tbh]
\centering
 \caption{\label{tab4}Network types and properties}
\begin{tabular}{|p{1cm}p{2cm}p{2cm}p{2cm}p{2cm}|}
    \hline
    \textbf{Type} & \textbf{M1} & \textbf{M2} & \textbf{M3}&\textbf{M4}\\
    \hline
    Type 1&Very low&Very low &Very low&Very low\\

    Type 2&Low/Medium&High&Low&Medium\\
    Type 3&Low/Medium&Medium&Low/Medium&High\\
    Type 4&Medium/High&Medium&Very high&Very high\\

    \hline
\end{tabular}
\end{table}
\subsection{Characteristics of Computer Science Venues}
The first question we address is that to what extend the venues in computer science are focused and how authors collaborate on that basic. In particular, we compare the properties of citation and co-authorship subgraph of journals and conferences to identify the differences between two types of publishing. To gain insights into the above questions, we process as follows: for all venues, we create their collaboration and citation subgraphs $G_a$ and $G_c$. Then we compute the four metrics defined in Section 9.1. For each metric we create a normalized histogram and by observing these histograms we are able to examine the characteristics of venues.

\begin{figure}
\centering

\includegraphics[height = 4in, width= 5in]{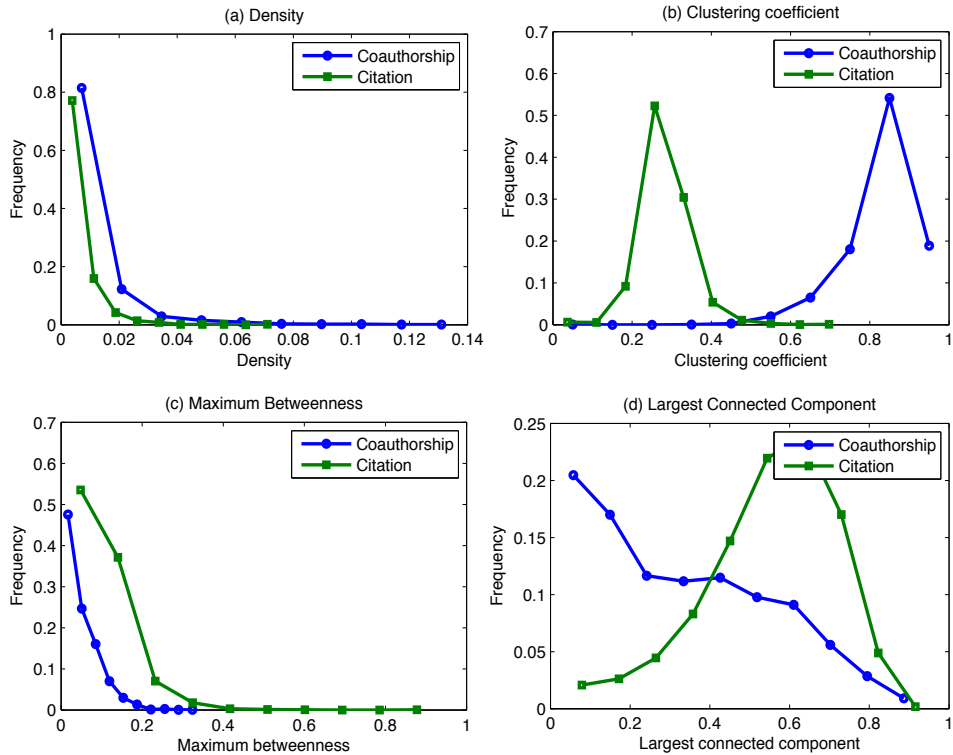}
\caption{Properties of collaboration and citation graphs of venues}
\label{fig:no10}
\end{figure}

The normalized histograms of the four metrics are given in Fig. \ref{fig:no10}. Firstly, most of the venues are not narrow, but they are indeed interdisciplinary (shown by low density and medium clustering coefficient of citation subgraphs in Fig. 10a and Fig. 10b). However, venues also tend to develop a main theme which is the main focused and closely related topics as the core topics. That is shown by big largest connected component in the citation subgraphs (Fig. 10d). According to the scheme in Table \ref{tab4}, most of the venues fall into the network type 3, characterized by low/medium density, medium clustering coefficient, low/medium maximum betweenness and big largest connected component.

We now consider the collaborative behavior of researchers in the venues. In Fig. \ref{fig:no10}, we can see that most of the co-authorship subgraphs are of network type 2 (low/medium density, high clustering coefficient, low maximum betweenness and medium largest connected component). That means researchers in the venues are clustered in disconnected working groups. The relative small number of venues that have big largest connected component (Fig. 10d) implies that though venues tend to develop the main theme, not so many of them successfully stimulates authors to collaborate on that theme. Low maximum betweenness (Fig. 10c) suggests that the gateways who connect several working groups rarely exist in the venues. We will investigate the relation between the existence of the gateways and the impact of venues in the next section.

\begin{figure}
\centering

\includegraphics[height = 4in, width= 5in]{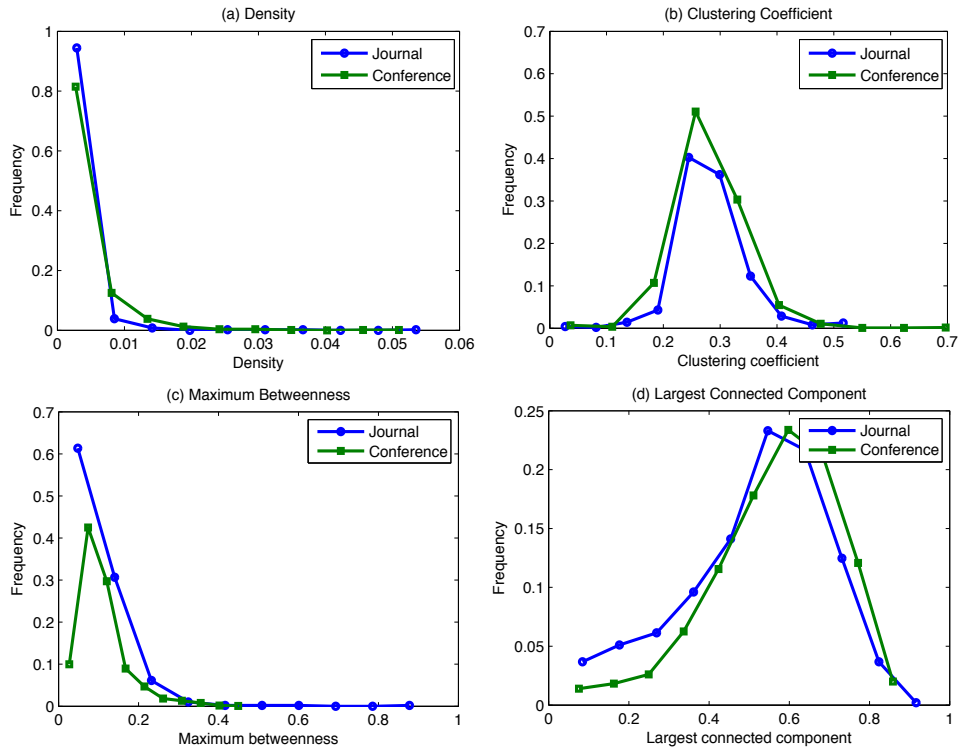}
\caption{A comparison of network properties of citation subgraph of journals and conferences}
\label{fig:no11}
\end{figure}

\begin{figure}
\centering

\includegraphics[height = 4in, width= 5in]{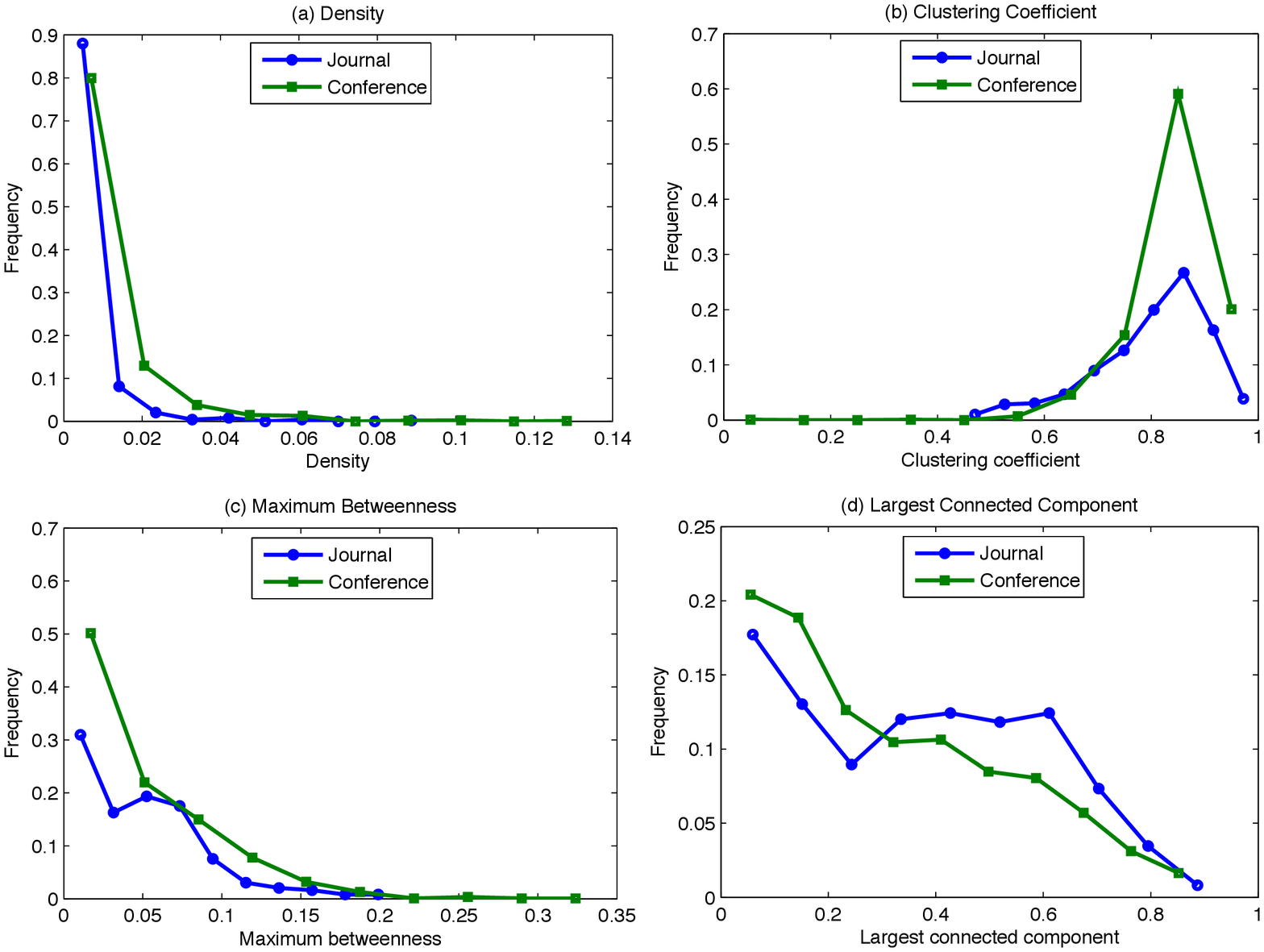}
\caption{A comparison of network properties of co-authorship subgraph of journals and conferences}
\label{fig:no12}
\end{figure}

Now we compare the properties of citation and co-authorship subgraphs of conference and journal. The question we try to address is that whether conferences expose the same pattern in citation and collaborative behavior as journals. Fig. \ref{fig:no11} and Fig. \ref{fig:no12} show the comparison of network properties of citation and co-authorship subgraphs of journal and conference. In general, most of journal and conference citation subgraphs are of network type 3 (low/medium density, medium clustering coefficient, medium maximum betweenness and big largest connected component). However, clustering coefficient of conferences' citation graph is higher than that of journals and maximum betweenness is lower. That means citations of conferences are placed in more disconnected clusters, which suggests that conferences are less focused than journals. A close look at the Fig. \ref{fig:no12} reveals some differences in collaborative behavior. Clustering coefficient and maximum betweenness of conferences' co-authorship subgraph are higher than journals', meaning that there exists more gateways in conferences than in journals and researchers in conferences tend to collaborate with peers in other working groups.

To summarize, venues in computer science are indeed interdisciplinary. Most of them established a core area while still connecting to other related areas. Journals are more focused than conferences, but conferences facilitate the communication between participants whose collaborations tend to cross different communities.

\subsection{Venues Subgraph and the PageRank}
\begin{figure}
\centering
\includegraphics[height = 4in, width= 5in]{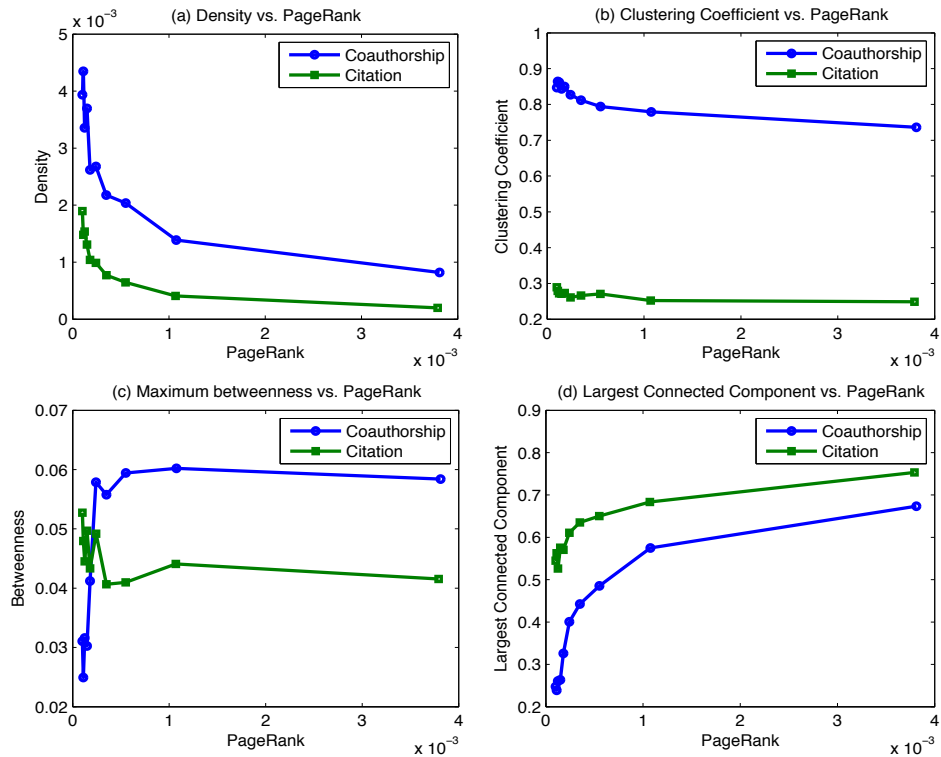}
\caption{Properties of collaboration and citation graphs of venues as a function of PageRank}
\label{fig:no13}
\end{figure}

Now we investigate the relation between the ranking of venues and the properties of collaboration and citation subgraphs. Our interest is whether the properties of collaboration and citation subgraphs reflect the impact of venues. In Fig. \ref{fig:no13}, we plot the median of the network properties for each PageRank value in order to analyze this relation.

Several observations can be made here. On the one hand, citation network of highly-ranked venues are of network type 3 (low/medium density, medium clustering coefficient, low/medium betweenness and big largest connected component), meaning that highly-ranked venues are focused. The co-authorship network of highly-ranked venues fall into network type 4, characterized by medium clustering coefficient, very high maximum betweenness and very big largest connected component. The vast majority of authors in those venues are connected in a large component and that component is connected to many other small groups via gatekeepers. On the other hand, it is not easy to identify the type of the citation subgraph of low-ranked venues. They might lay between network type 2 and type 3, with high/medium clustering coefficient, low/medium maximum betwenness and low/medium largest connected component. However, co-authorship subgraph of low-ranked venues are clearly of type 2, where authors are clustered in disconnected groups.

To summarize, highly-ranked venues are focused as they develop the main topics as the core and successfully motivate authors to collaborate on these topics. In these venues, there exist key members who connect different subgroups to the core. They serve as a gate to join the new ideas to the main theme of the venue. This is very important for every community of practice since one of the key success factors is not only to retain the well-developed ideas but also attract people to bring new ideas to the community \cite{Kien05,Kienle05,Kienle06}. Although low ranked venues might also develop the main theme, but they mostly do not successfully build up a large community to work on that or they are still in the early phase of developing their community.

\section{Conclusions}
In this paper, we presented our study on knowledge network of computer science. Based on the combined DBLP and CiteSeerX databases, the knowledge network is generated using both journal and conference publications. The visualizations show the cluster structure of computer science knowledge network, which is not possible by the analysis of journals-only. Venues of the same fields or related fields are grouped into clusters which can be defined as disciplines or sub-disciplines. We analyze the development of computer science disciplines by visualizing the knowledge network at different time points. One important conclusion is that conferences constitute the social structures that shape the computer science knowledge and the field is becoming more interdisciplinary as sub-disciplines are connected to many others.

We analyze the citation and collaboration subgraphs of venues by different SNA metrics. We find that venues are interdisciplinary and they develop their core topics as the main focus. By comparing the citation, collaboration subgraphs of journal and conference, our study shows that though journals are more focused than conferences, the latter facilitate the communication between researchers. We further analyze the relation between the impact and the properties of citation and collaboration subgraph of venues. One important conclusion is that highly ranked venues successfully develop their theme as well as their community and experts are the key success factor for the development of a venue. That confirms one of the principle for cultivating scientific community of practice studied by several researchers.

In the future, more digital libraries need to be integrated to obtain complete citation information. Given the objective of this paper is to study the macro structure of computer science, DBLP and CiteSeerX are quite sufficient. However, to study the structure of knowledge network at more detail and local level (i.e at the sub-discipline level), more citation data and venue proceedings are needed.  Several datasets are possible, e.g ACM, IEEE Xplore, Microsoft Academic Research, CEUR-WS.org\footnote{\url{http://sunsite.informatik.rwth-aachen.de/Publications/CEUR-WS/}}. Citation information could also be gathered from search engines like Google Scholar. Furthermore, the ranking studied in this paper is global ranking. It probably does not reflect the complete importance of a venue in a particular field, especially in some more marginal disciplines such as computer graphics or bioinformatics. Therefore a deep analysis and ranking at the sub-discipline level is necessary to gain an insight into a particular domain and to have a full evaluation of the impact of venues.

\section*{Acknowledgment}
This work has been supported by the Graduiertenkolleg (GK) ``Software for mobile communication
system'', RWTH Aachen University, the BIT Research School by RWTH Aachen University and the University of Bonn, and the EU FP7 IP ROLE. We would like to thank our colleagues for the fruitful discussions.
%-------------------------------------------------------------------------

%-------------------------------------------------------------------------
%\nocite{ex1,ex2}
%\bibliographystyle{latex8}
\bibliographystyle{spmpsci}
\bibliography{SNAM_Revision}

\begin{thebibliography}{10}
\providecommand{\url}[1]{{#1}}
\providecommand{\urlprefix}{URL }
\expandafter\ifx\csname urlstyle\endcsname\relax
  \providecommand{\doi}[1]{DOI~\discretionary{}{}{}#1}\else
  \providecommand{\doi}{DOI~\discretionary{}{}{}\begingroup
  \urlstyle{rm}\Url}\fi

\bibitem{Huang06}
0002, J.H., Ertekin, S., Giles, C.L.: Efficient name disambiguation for
  large-scale databases.
\newblock In: PKDD, pp. 536--544 (2006)

\bibitem{Bollen09}
Bollen, J., Van~de Sompel, H., Hagberg, A., Bettencourt, L., Chute, R.,
  Rodriguez, M.A., Balakireva, L.: Clickstream data yields high-resolution maps
  of science.
\newblock PLoS ONE \textbf{4}(3), e4803+ (2009).
\newblock \doi{10.1371/journal.pone.0004803}.
\newblock \urlprefix\url{http://dx.doi.org/10.1371/journal.pone.0004803}

\bibitem{Bollen092}
Bollen, J., de~Sompel, H.V., Hagberg, A.A., Chute, R.: A principal component
  analysis of 39 scientific impact measures.
\newblock CoRR \textbf{abs/0902.2183} (2009)

\bibitem{Boyack07}
Boyack, K.W., Börner, K., Klavans, R.: Mapping the structure and evolution of
  chemistry research.
\newblock In: In D. Torres-Salinas and H. Moed (Eds.), Proceedings of the 11th
  International Conference of Scientometrics and Informetrics, p. 112123 (2007)

\bibitem{Boyack05}
Boyack, K.W., Klavans, R., B\"{o}rner, K.: Mapping the backbone of science.
\newblock Scientometrics \textbf{64}(3), 351--374 (2005).
\newblock \doi{10.1007/s11192-005-0255-6}.
\newblock \urlprefix\url{http://dx.doi.org/10.1007/s11192-005-0255-6}

\bibitem{Brin98}
Brin, S., Page, L.: The anatomy of a large-scale hypertextual web search
  engine.
\newblock Comput. Netw. ISDN Syst. \textbf{30}(1-7), 107--117 (1998).
\newblock \doi{http://dx.doi.org/10.1016/S0169-7552(98)00110-X}

\bibitem{Chen10}
Chen, J., Konstan, J.A.: Conference paper selectivity and impact.
\newblock Commun. ACM \textbf{53}, 79--83 (2009).
\newblock \doi{http://doi.acm.org/10.1145/1743546.1743569}.
\newblock \urlprefix\url{http://doi.acm.org/10.1145/1743546.1743569}

\bibitem{clauset-2004-70}
Clauset, A., Newman, M.E.J., Moore, C.: Finding community structure in very
  large networks.
\newblock Physical Review E \textbf{70}, 066,111 (2004).
\newblock \urlprefix\url{doi:10.1103/PhysRevE.70.066111}

\bibitem{Coleman88}
Coleman, J.S.: Social capital in the creation of human capital.
\newblock The American Journal of Sociology \textbf{94}, pp. S95--S120 (1988).
\newblock \urlprefix\url{http://www.jstor.org/stable/2780243}

\bibitem{Ding00}
Ding, Y., Chowdhury, G., Foo, S.: Journal as markers of intellectual space:
  Journal co-citation analysis of information retrieval area.
\newblock Scientometrics \textbf{47}(1), 55--73 (2000)

\bibitem{Fortnow09}
Fortnow, L.: Viewpoint: Time for computer science to grow up.
\newblock Commun. ACM \textbf{52}, 33--35 (2009).
\newblock \doi{http://doi.acm.org/10.1145/1536616.1536631}.
\newblock \urlprefix\url{http://doi.acm.org/10.1145/1536616.1536631}

\bibitem{Freyne10}
Freyne, J., Coyle, L., Smyth, B., Cunningham, P.: Relative status of journal
  and conference publications in computer science.
\newblock Commun. ACM \textbf{53}, 124--132 (2010).
\newblock \doi{http://doi.acm.org/10.1145/1839676.1839701}.
\newblock \urlprefix\url{http://doi.acm.org/10.1145/1839676.1839701}

\bibitem{fruchterman91}
Fruchterman, T.M.J., Reingold, E.M.: Graph drawing by force-directed placement.
\newblock Softw. Pract. Exper. \textbf{21}(11), 1129--1164 (1991).
\newblock \doi{http://dx.doi.org/10.1002/spe.4380211102}

\bibitem{Frederic}
Gilbert, F., Simonetto, P., Zaidi, F., Jourdan, F., Bourqui, R.: Communities
  and hierarchical structures in dynamic social networks: analysis and
  visualization.
\newblock Social Network Analysis and Mining pp. 1--13 (2010).
\newblock \urlprefix\url{http://dx.doi.org/10.1007/s13278-010-0002-8}.
\newblock 10.1007/s13278-010-0002-8

\bibitem{granovetter83}
Granovetter, M.: {The Strength of Weak Ties: A Network Theory Revisited}.
\newblock Sociological Theory \textbf{1}, 201--233 (1983)

\bibitem{Han04}
Han, H., Giles, C.L., Zha, H., Li, C., Tsioutsiouliklis, K.: Two supervised
  learning approaches for name disambiguation in author citations.
\newblock In: JCDL, pp. 296--305 (2004)

\bibitem{Hirsch05}
Hirsch, J.E.: {An index to quantify an individual's scientific research
  output}.
\newblock Proceedings of the National Academy of Sciences \textbf{102}(46),
  16,569--16,572 (2005).
\newblock \doi{10.1073/pnas.0507655102}.
\newblock \urlprefix\url{http://dx.doi.org/10.1073/pnas.0507655102}

\bibitem{Kienle05}
Kienle, A., Wessner, M.: Our way to taipei: an analysis of the first ten years
  of the cscl community.
\newblock In: Proceedings of th 2005 conference on Computer support for
  collaborative learning: learning 2005: the next 10 years!, CSCL '05, pp.
  262--271. International Society of the Learning Sciences (2005).
\newblock \urlprefix\url{http://portal.acm.org/citation.cfm?id=1149293.1149327}

\bibitem{Kien05}
Kienle, A., Wessner, M.: Principles for cultivating scientific communities of
  practice.
\newblock In: P.~Besselaar, G.~Michelis, J.~Preece, C.~Simone (eds.)
  Communities and Technologies 2005, pp. 283--299. Springer Netherlands (2005).
\newblock \urlprefix\url{http://dx.doi.org/10.1007/1-4020-3591-8_15}

\bibitem{Kienle06}
Kienle, A., Wessner, M.: Analysing and cultivating scientific communities of
  practice.
\newblock Int. J. Web Based Communities \textbf{2}, 377--393 (2006).
\newblock \doi{10.1504/IJWBC.2006.011765}.
\newblock \urlprefix\url{http://portal.acm.org/citation.cfm?id=1359102.1359103}

\bibitem{Klavans06}
Klavans, R., Boyack, K.W.: Identifying a better measure of relatedness for
  mapping science.
\newblock J. Am. Soc. Inf. Sci. Technol. \textbf{57}(2), 251--263 (2006).
\newblock \doi{http://dx.doi.org/10.1002/asi.v57}

\bibitem{Kumar09}
Kumar, M.: {Evaluating Scientists: Citations, Impact Factor, h-Index, Online
  Page Hits and What Else?}
\newblock IETE Technical Review \textbf{26}(3), 165--168 (2009).
\newblock \doi{10.4103/0256-4602.50699}

\bibitem{Lambiotte09}
Lambiotte, R., Panzarasa, P.: Communities, knowledge creation, and information
  diffusion.
\newblock Journal of Informetrics \textbf{3}(3), 180 -- 190 (2009).
\newblock \doi{DOI: 10.1016/j.joi.2009.03.007}.
\newblock
  \urlprefix\url{http://www.sciencedirect.com/science/article/B83WV-4W8TJP2-3/%
2/c82036ec7db02965d6ff85342d55236c}.
\newblock Science of Science: Conceptualizations and Models of Science

\bibitem{Lee05}
Lee, D., On, B.W., Kang, J., Park, S.: Effective and scalable solutions for
  mixed and split citation problems in digital libraries.
\newblock In: IQIS, pp. 69--76 (2005)

\bibitem{Ley09}
Ley, M.: Dblp - some lessons learned.
\newblock PVLDB \textbf{2}(2), 1493--1500 (2009)

\bibitem{leydesdorff04}
Leydesdorff, L.: Clusters and maps of science journals based on bi-connected
  graphs in the journal citation reports.
\newblock Journal of Documentation \textbf{60}(4), 317 (2004).
\newblock
  \urlprefix\url{http://www.citebase.org/abstract?id=oai:arXiv.org:0912.1221}

\bibitem{Leydesdorff042}
Leydesdorff, L.: Top-down decomposition of the journal citation reportof the
  social science citation index: Graph- and factor-analytical approaches.
\newblock Scientometrics \textbf{60}(2), 317 (2004).
\newblock
  \urlprefix\url{http://www.citebase.org/abstract?id=oai:arXiv.org:0912.1221}

\bibitem{Leydesdorff06}
Leydesdorff, L.: Betweenness centrality as an indicator of the
  interdisciplinarity of scientific journals.
\newblock J. Am. Soc. Inf. Sci. Technol. \textbf{58}(9), 1303--1319 (2007).
\newblock \doi{http://dx.doi.org/10.1002/asi.20614}

\bibitem{Dalhia11}
Mani, D., Knoke, D.: On intersecting ground: the changing structure of us
  corporate networks.
\newblock Social Network Analysis and Mining \textbf{1}, 43--58 (2011).
\newblock \urlprefix\url{http://dx.doi.org/10.1007/s13278-010-0013-5}.
\newblock 10.1007/s13278-010-0013-5

\bibitem{McCain98}
McCain, K.W.: Neural networks research in context: A longitudinal journal
  cocitation analysis of an emerging interdisciplinary field.
\newblock Scientometrics \textbf{5}(5), 389--410 (1998)

\bibitem{MacCallum00}
McCallum, A., Nigam, K., Ungar, L.H.: Efficient clustering of high-dimensional
  data sets with application to reference matching.
\newblock In: KDD '00: Proceedings of the sixth ACM SIGKDD International
  Conference on Knowledge Discovery and Data Mining, pp. 169--178. ACM, New
  York, NY, USA (2000).
\newblock \doi{http://doi.acm.org/10.1145/347090.347123}

\bibitem{Meyer09}
Meyer, B., Choppy, C., Staunstrup, J., van Leeuwen, J.: Viewpoint: Research
  evaluation for computer science.
\newblock Commun. ACM \textbf{52}, 31--34 (2009).
\newblock \doi{http://doi.acm.org/10.1145/1498765.1498780}.
\newblock \urlprefix\url{http://doi.acm.org/10.1145/1498765.1498780}

\bibitem{Morris98}
Morris, T.A., McCain, K.W.: The structure of medical informatics journal
  literature.
\newblock Journal of the American Medical Informatics Association
  \textbf{5}(5), 448--566 (1998)

\bibitem{Moya04}
Moya-Aneg\'on, F., Vargas-Quesada, B., Herrero-Solana, V.,
  Chinchilla-Rodr\'iguez, Z., Corera-\'Alvarez, E., noz Fern\'andez, F.J.M.: A
  new technique for building maps of large scientific domains based on the
  co-citation of classes and categories.
\newblock Scientometrics \textbf{61}(1), 129--145 (2004).
\newblock \urlprefix\url{http://eprints.rclis.org/10899/}

\bibitem{newman-2004-692}
Newman, M.E.J.: Fast algorithm for detecting community structure in networks.
\newblock Physical Review E \textbf{69}, 066,133 (2004).
\newblock
  \urlprefix\url{http://www.citebase.org/abstract?id=oai:arXiv.org:cond-mat/03%
09508}

\bibitem{newman-2006-103}
Newman, M.E.J.: Modularity and community structure in networks.
\newblock Proc.Natl.Acad.Sci.USA \textbf{103}, 8577 (2006).
\newblock \urlprefix\url{doi:10.1073/pnas.0601602103}

\bibitem{newman-2004-69}
Newman, M.E.J., Girvan, M.: Finding and evaluating community structure in
  networks.
\newblock Physical Review E \textbf{69}, 026,113 (2004).
\newblock
  \urlprefix\url{http://www.citebase.org/abstract?id=oai:arXiv.org:cond-mat/03%
08217}

\bibitem{Pereira09}
Pereira, D.A., Ribeiro-Neto, B.A., Ziviani, N., Laender, A.H.F., Gon\c{c}alves,
  M.A., Ferreira, A.A.: Using web information for author name disambiguation.
\newblock In: JCDL, pp. 49--58 (2009)

\bibitem{Pham10}
Pham, M., Klamma, R.: The structure of the computer science knowledge network.
\newblock In: 2010 International Conference on Advances in Social Networks
  Analysis and Mining (ASONAM), pp. 17 --24 (2010).
\newblock \doi{10.1109/ASONAM.2010.58}

\bibitem{Seglen97}
Seglen, P.O.: Why the impact factor of journals should not be used for
  evaluating research.
\newblock BMJ: British Medical Journal \textbf{314}, 498--502 (1997)

\bibitem{Shi10}
Shi, X., Leskovec, J., McFarland, D.A.: Citing for high impact.
\newblock In: Proceedings of the 10th annual joint conference on Digital
  libraries, JCDL '10, pp. 49--58. ACM, New York, NY, USA (2010).
\newblock \doi{http://doi.acm.org/10.1145/1816123.1816131}.
\newblock \urlprefix\url{http://doi.acm.org/10.1145/1816123.1816131}

\bibitem{CACMStaff09}
Staff, C.: Pay for editorial independence.
\newblock Commun. ACM \textbf{52}, 6--7 (2009).
\newblock \doi{http://doi.acm.org/10.1145/1592761.1592764}.
\newblock \urlprefix\url{http://doi.acm.org/10.1145/1592761.1592764}

\bibitem{Treeratpituk09}
Treeratpituk, P., Giles, C.L.: Disambiguating authors in academic publications
  using random forests.
\newblock In: JCDL, pp. 39--48 (2009)

\bibitem{Tsay03}
Tsay, M.Y., Xu, H., wen Wu, C.: Journal co-citation analysis of semiconductor
  literature.
\newblock Scientometrics \textbf{57}(1), 7--25 (2003)

\bibitem{Vardi09}
Vardi, M.Y.: Conferences vs. journals in computing research.
\newblock Commun. ACM \textbf{52}, 5--5 (2009).
\newblock \doi{http://doi.acm.org/10.1145/1506409.1506410}.
\newblock \urlprefix\url{http://doi.acm.org/10.1145/1506409.1506410}

\bibitem{Wasserman95}
Wasserman, S., Faust, K.: Social Network Analysis: Methods and Applications
  (Structural Analysis in the Social Sciences).
\newblock {Cambridge University Press} (1995).
\newblock
  \urlprefix\url{http://www.amazon.com/exec/obidos/redirect?tag=citeulike07-20%
&path=ASIN/0521382696}

\bibitem{Leyla10}
Zhuhadar, L., Nasraoui, O., Wyatt, R., Yang, R.: Visual knowledge
  representation of conceptual semantic networks.
\newblock Social Network Analysis and Mining pp. 1--11 (2010).
\newblock \urlprefix\url{http://dx.doi.org/10.1007/s13278-010-0008-2}.
\newblock 10.1007/s13278-010-0008-2

\end{thebibliography}

\end{document}